%% file: main.tex
\documentclass[10pt,twocolumn,letterpaper]{article}

\usepackage{cvpr}              

\usepackage{graphicx}
\usepackage{amsmath}
\usepackage{amssymb}
\usepackage{booktabs}
\usepackage{enumitem}
\usepackage{tikz-cd}
\usepackage{enumitem}
\usepackage{siunitx}
\usepackage{stackengine} 
\usepackage{mathtools}
\usepackage{amsmath}
\usepackage{bm}
\usepackage{amsthm}
\usepackage{appendix}
\usepackage[accsupp]{axessibility}
\usepackage{xr}
\newcommand{\norm}[1]{\lVert#1\rVert}
\newcommand{\RN}[1]{%
  \textup{\uppercase\expandafter{\romannumeral#1}}%
}
\newcommand{\myvspace}{\vspace{-3pt}}

\newtheorem{definition}{Definition}
\newtheorem{lemma}{Lemma}

\newtheorem{example}{Example}

\DeclarePairedDelimiterX{\infdivx}[2]{(}{)}{%
  #1\;\delimsize\|\;#2%
  }
\newcommand{\infdiv}{\text{KL}\infdivx}
\usepackage[pagebackref,breaklinks,colorlinks]{hyperref}

\usepackage[capitalize]{cleveref}
\crefname{section}{Sec.}{Secs.}
\Crefname{section}{Section}{Sections}
\Crefname{table}{Table}{Tables}
\crefname{table}{Tab.}{Tabs.}

\def\cvprPaperID{10} 
\def\confName{CVPR}
\def\confYear{2023}

\makeatletter
\providecommand{\@EveryShipout@Hook}{}
\makeatother

\begin{document}

\title{Quantifying Extrinsic Curvature in Neural Manifolds}

\author{Francisco Acosta\textsuperscript{1},
 Sophia Sanborn\textsuperscript{2},
 Khanh Dao Duc\textsuperscript{3},
 Manu Madhav\textsuperscript{4},
 Nina Miolane\textsuperscript{2}\\
\textsuperscript{1} Physics, \textsuperscript{2} Electrical and Computer Engineering, UC Santa Barbara\\
\textsuperscript{3} Mathematics, \textsuperscript{4} Biomedical Engineering, University of British Columbia\\
{\tt\small {facosta@ucsb.edu}, {sanborn@ucsb.edu}, {kdd@math.ubc.ca}}\\ 
{\tt\small {manu.madhav@ubc.ca}, {ninamiolane@ucsb.edu}}
}

\maketitle

\begin{abstract}
The neural manifold hypothesis postulates that the activity of a neural population forms a low-dimensional manifold whose structure reflects that of the encoded task variables. 
In this work, we combine topological deep generative models and extrinsic Riemannian geometry to introduce a novel approach for studying the structure of neural manifolds. This approach (i) computes an explicit parameterization of the manifolds and (ii) estimates their local extrinsic curvature\textemdash hence quantifying their shape within the neural state space. Importantly, we prove that our methodology is invariant with respect to transformations that do not bear meaningful neuroscience information, such as permutation of the order in which neurons are recorded. We show empirically that we correctly estimate the geometry of synthetic manifolds generated from smooth deformations of circles, spheres, and tori, using realistic noise levels. We additionally validate our methodology on simulated and real neural data, and show that we recover geometric structure known to exist in hippocampal place cells. We expect this approach to open new avenues of inquiry into geometric neural correlates of perception and behavior.
\end{abstract}

\section{Introduction}
\label{sec:intro}
A fundamental idea in machine learning is the \textit{manifold hypothesis}~\cite{bengio2013representation}, which postulates that many kinds of real-world data occupy a lower-dimensional manifold embedded in the high-dimensional data space. This perspective has also informed the analysis of neural population activity, where it is hypothesized that neural representations form low-dimensional manifolds whose structure reflects the structure of the task variables they encode\textemdash the \textit{neural manifold hypothesis}~\cite{dicarlo2007untangling,gao2015simplicity,chung2021neural, vyas2020computation}.

A clear example of this can be observed in the brain's so-called ``cognitive maps'' of space, encoded in the activities of place cells in the hippocampus~\cite{okeefe1971hippocampus} and grid cells in the entorhinal cortex~\cite{hafting2005microstructure}. These neural populations are responsible for tracking an animal's position as it navigates physical space. Grid cells exhibit striking regularity in their firing patterns, tiling space in hexagonal grids \cite{moser2008place}. It has been observed, however, that these grid maps of space do not always faithfully reflect physical distances. Rather, the animal's cognitive map can be warped and distorted by the reward or relevance associated with different regions of the map \cite{boccara2019entorhinal, keinath2018environmental, savelli2017framing}. Here, we propose a method that can explicitly quantify and analyze this kind of neural warping.

\begin{figure}
    \centering
    \vspace{-0.2cm}
    \includegraphics[width=0.9\linewidth]{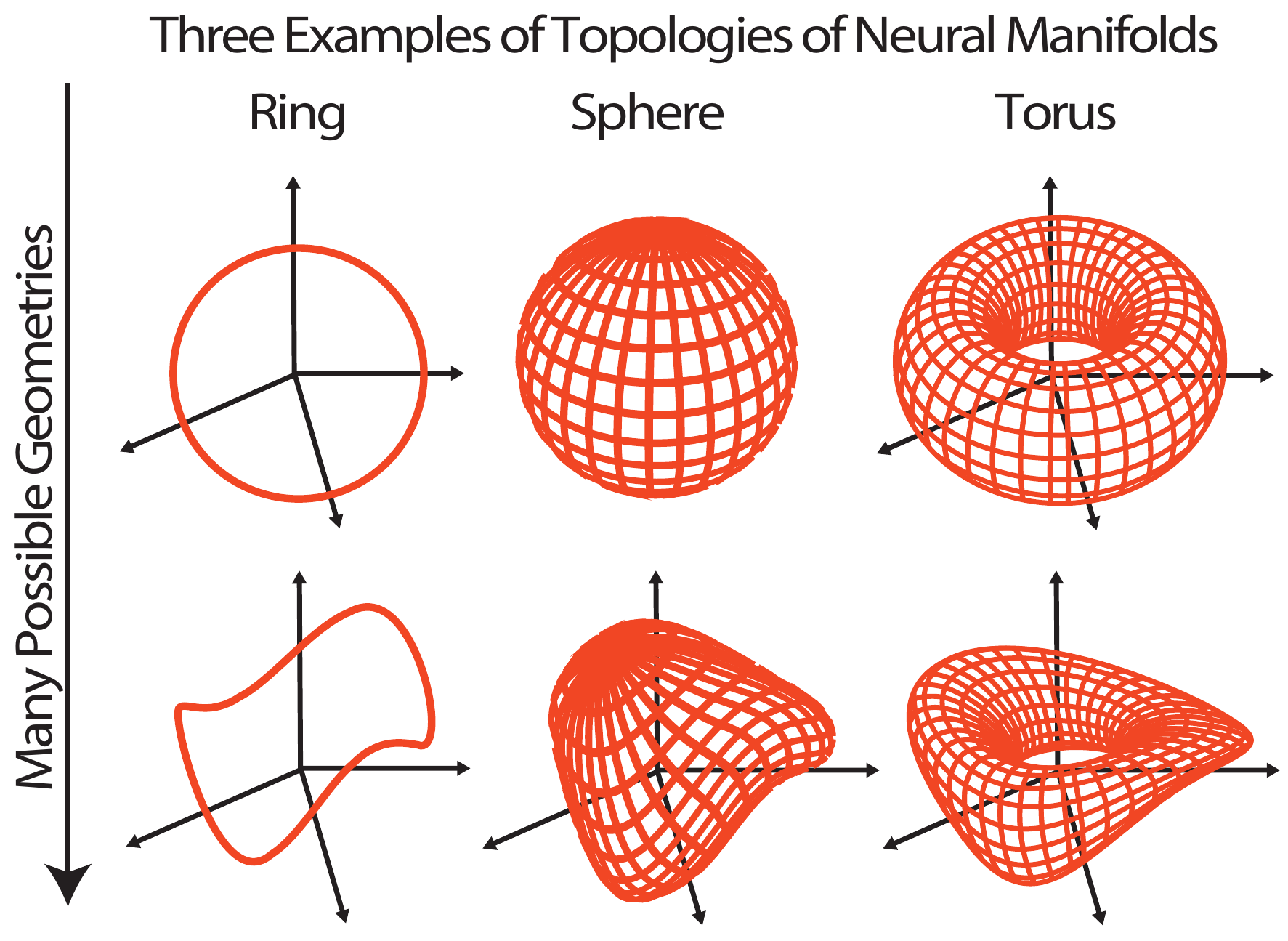}
    \caption{Topology versus geometry. Topological methods reveal whether the neural manifold is a ring, a sphere, or a torus, i.e., whether it belongs to one of the three columns. Our geometric analysis determines the curvature of the ring, sphere, or torus \textemdash hence revealing their \textit{neural shape}.}
    \label{fig:topo-versus-geom}
\end{figure}

We introduce a novel approach to reveal the geometry of neural manifolds. When applied to high-dimensional point-cloud data, our method (i) computes an explicit parameterization of the underlying manifold and (ii) estimates its local \textit{extrinsic} curvature, hence providing a concrete quantification of its neural shape within the neural state space (see Fig.~\ref{fig:topo-versus-geom}). Our contributions are as follows.

\begin{enumerate}[leftmargin=*]
    \item We present a Riemannian method for estimating the extrinsic curvatures of neural manifolds, which leverages topological Variational Autoencoders (VAEs). We provide intuition for the mathematics behind this approach with the analytical calculation of curvature for three manifolds commonly encountered in neuroscience: rings, spheres, and tori (Fig.~\ref{fig:topo-versus-geom}).\myvspace
    \item We demonstrate that this method is invariant under (a) reparameterization of the VAE's latent space and (b) permutation of the order in which recorded neurons appear in the data array. As such, our approach is appropriate for recovering meaningful geometric structure in real-world neuroscience experiments.\myvspace
    \item We quantify the performance of the method applied to synthetic manifolds with known curvature profiles, by computing a curvature estimation error under varied numbers of recorded neurons and simulated, yet realistic, measurement noise conditions. We further demonstrate its application to simulated place cell data\textemdash cells from neural circuits involved in navigation.\myvspace
    \item We successfully apply our method to real hippocampal place cells recorded from rodents moving along a circular track~\cite{jayakumar2019recalibration,madhav2022dome}. We show that we reveal geometric structure that is consistent with results on simulations.\myvspace
\end{enumerate}

This paper thus proposes a first-of-its-kind approach for explicitly quantifying the extrinsic curvature of a neural manifold. Our goal is to provide the neuroscience community with tools to rigorously parameterize and quantify the geometry of neural representations. All code is publicly available and new differential geometric tools have been incorporated in the open-source software \texttt{Geomstats}~\cite{miolane2020geomstats}.

\section{Related Works}


\paragraph{Learning Low-Dimensional Structure of Neural Manifolds} Many approaches to uncovering manifold structure in neural population activity rely on dimensionality reduction techniques such as Principal Component Analysis (PCA), Isomap, Locally Linear Embedding (LLE), and t-Distributed Stochastic Neighbor Embedding (t-SNE) ~\cite{pearson1901lines, tenenbaum2000global, roweis2000nonlinear, vandermaaten2008visualizing}. While these techniques can reveal the existence of lower-dimensional structure in neural population activity, they do not provide an explicit parameterization of the neural manifold, and often misrepresent manifolds with non-trivial topology \cite{wattenberg2016use}.

\paragraph{Learning the Topology of Neural Manifolds} Methods based on topological data analysis (TDA), such as persistent homology, have begun to reveal \textit{topological} structure in neural representations and the task variables they encode~\cite{gardner2022toroidal,chaudhuri2019intrinsic,curto2017can, Hermansen2022}. Topological methods can identify when two manifolds have the same number of holes\textemdash differentiating, for example, a ring from a sphere or a torus (see topologies in Fig.~\ref{fig:topo-versus-geom}). While these methods are able to uncover properties of the manifold that are invariant to continuous deformations, they do not necessarily capture the strictly \textit{geometric} properties of the manifold\textemdash \textit{i.e.}, structure that \textit{changes} under continuous deformations, such as curvature (see vertical axis in Fig.~\ref{fig:topo-versus-geom}). 
To date, few methods exist to explicitly quantify and parameterize the geometric structure of neural manifolds. A geometric parameterization of neural manifolds would permit a more precise understanding of the behavioral or perceptual relationships between points in the neural state space.

\paragraph{Learning Riemannian Geometry with Deep Generative Models} Recent theoretical advancements have permitted the analysis of the Riemannian geometry of manifolds learned in deep neural network models, including VAEs \cite{hauser2017principles, shao2018riemannian, chadebec2020geometry,connor2021variational}. These analyses permit geometry-aware statistics on the model latent space \cite{kuhnel2018latent,chen2018metrics}, and can be used to improve training~\cite{arvanitidis2017latent,kalatzis2020variational,chadebec2022geometric}. In addition to its application in the neuroscience domain, our work also extends these techniques. Thus far, all such approaches have focused exclusively on \textit{intrinsic} notions of curvature, as given by the Riemann curvature tensor and contractions thereof such as the Ricci tensor or the scalar curvature tensor~\cite{hauser2017principles,shao2018riemannian,kuhnel2018latent}. We extend these methods with an approach capable of quantifying the \textit{extrinsic} curvature of latent manifolds, which\textemdash as we will argue\textemdash is more suitable to describe manifolds emerging in experimental neuroscience.


\begin{figure*}
    \centering
    \includegraphics[width=\textwidth]{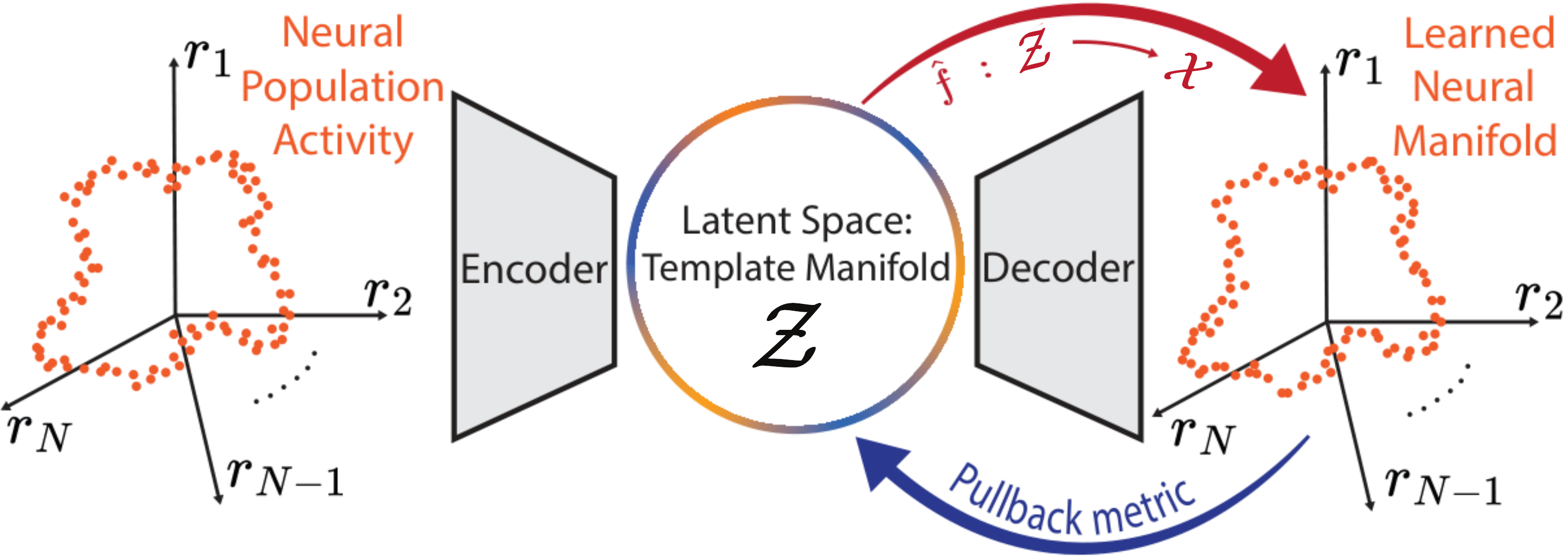}
    \caption{\textbf{Methods Overview.} Neural activity vectors are represented as orange points in $\mathcal{X} =\mathbb{R}_{+}^N$: they correspond either to the activation of an artificial neural network layer with $N$ neurons, or to the electrical recordings of $N$ biological neurons. Together, the set of neural activity vectors forms a neural manifold. Our method uses a topologically-aware variational autoencoder (VAE), whose latent space has the (known) topology of the neural manifold, shown here as the ring $\mathcal{S}^1$. The decoder provides an estimate of the differentiable function $\hat{f}$ whose derivatives yield the Riemannian metric, as well as the intrinsic and extrinsic curvatures of the neural manifold.}
    \label{fig:summary}
\end{figure*}

\section{A Riemannian Approach to Neural Population Geometry}\label{sec:methods}

We propose a method to compute geometric quantities\textemdash the \textit{first fundamental form} (i.e.\ the \textit{pullback metric}), the \textit{second fundamental form}, and the \textit{mean curvature vector} at every point on a neural manifold\textemdash resulting in a powerful and precise description of the geometry of neural population activity. 


\subsection{Overview}

We denote with $\mathcal{M}$ the neural manifold of interest, which represents the population activity of $N$ neurons in the neural state space $\mathcal{X} = \mathbb{R}_{+}^N$ as illustrated in Fig.~\ref{fig:topo-versus-geom} and Fig.~\ref{fig:summary}. Our approach learns the neural population geometry in three steps.

\paragraph{1. Learn the Topology:} 
The topology of $\mathcal{M}$ can be determined using topological data analysis methods such as persistent homology~\cite{ghrist2007barcodes}. We define the template manifold $\mathcal{Z}$ as the canonical manifold homeomorphic to $\mathcal{M}$. For example, for a neural manifold with the topology of a one-dimensional ring, we choose $\mathcal{Z}$ to be the circle, $\mathcal{S}^{1}$. In our neuroscience applications, we specifically consider template manifolds that are either $n$-spheres $\mathcal{S}^n$, or direct products of $n$-spheres, as these include the most common topological manifolds observed in neuroscience experiments:  $\mathcal{S}^1$, $\mathcal{S}^2$ and $\mathcal{T}^2 = \mathcal{S}^1 \times \mathcal{S}^1$.

\paragraph{2. Learn the Deformation:}
 We determine the mapping $f: \mathcal{Z} \to \mathcal{X}$ that characterizes the smooth deformation from the template $\mathcal{Z}$ to $\mathcal{X}$. We encode $f$ with a neural network and propose to learn it with a variational autoencoder (VAE)~\cite{kingma2013auto} trained as a latent variable model of neural activity in $\mathcal{X} = \mathbb{R}_{+}^N$ for $N$ neurons. The VAE's latent space is topologically-constrained to be $\mathcal{Z}$, as in ~\cite{falorsi2018explorations, davidson2018hyperspherical, mikulski2019toroidal, mathieu2019continuous}. After training, the VAE's decoder provides an estimate $\hat{f}$ of $f$, i.e.\ a differentiable function whose derivatives yield the Riemannian metric, and intrinsic and extrinsic curvatures of the neural manifold \textemdash see Fig.\ref{fig:summary}.

\paragraph{3. Learn the Geometry:} By using a decoder that is twice differentiable, via neural network activation functions such as tanh$(\cdot)$ and softplus$(\cdot)$, and considering the fact that invertible matrices are a dense subset in the space of square matrices, the map $\hat{f}$ is a diffeomorphism from $\mathcal{Z}$ to the (learned) neural manifold $\hat{\mathcal{M}}$, and an immersion of $\mathcal{Z}$ into the neural state space $\mathcal{X} = \mathbb{R}_{+}^N$.
As such, $\hat{f}$ allows us to endow a pullback metric on the latent space template manifold $\mathcal{Z}$, which characterizes the Riemannian geometry of $\mathcal{M}$ embedded in the neural state space $\mathcal{X}$.
We use automatic differentiation to compute the pullback metric and curvatures from $\hat{f}$.

\subsection{Learning the Deformation with Topologically-Aware VAEs}\label{sec:our-vae}

Variational Autoencoders (VAEs)\cite{kingma2013auto} are probabilistic deep generative models that learn to compress data into a latent variable, revealing latent manifold structure in the process. The supplementary materials provide an introduction to this framework. In a standard VAE, latent variables take values in Euclidean space, $z \in \mathbb{R}^L$ (where typically $L < N$), and their prior distribution $p(z)$ is assumed to be Gaussian with unit variance, $p(z) = \mathcal{N}(0,\mathbb{I}_{L})$. While these assumptions are mathematically convenient, they are not suitable for modeling data whose latent variables lie on manifolds with nontrivial topology~\cite{davidson2018hyperspherical,falorsi2018explorations}. Our approach instead constrains the latent space of the VAE to the template manifold $\mathcal{Z}$, assumed to be an $n$-sphere $\mathcal{S}^n$ or direct products thereof. We follow the implementation of a hyperspherical VAE ~\cite{davidson2018hyperspherical} and a toroidal VAE~\cite{bjerke2022understanding} to accommodate the product of circles as well. 



\paragraph{Hyperspherical VAEs} The hyperspherical VAE uses the uniform distribution on the $n$-sphere $\text{U}(\mathcal{S}^n)$ as its non-informative prior $p(z)$, and a von Mises-Fisher $\text{vMF}(\mu, \kappa)$ distribution as its approximate posterior.
We follow \cite{davidson2018hyperspherical} to implement the regularization loss corresponding to the Kullback-Leibler divergence for the hyperspherical VAE and also use their proposed sampling procedure for $\text{vMF}$, which gives a reparameterization trick appropriate for these distributions. Hyperspheres represent a very important class of neural manifolds, e.g.\ emerging in the circular structure of the head direction circuit~\cite{chaudhuri2019intrinsic}.

\paragraph{Toroidal VAEs}
To accommodate additional neural manifolds of interest to neuroscience, we extend the hyperspherical VAE by implementing a toroidal VAE: $\mathcal{T}^n$-VAE~\cite{bjerke2022understanding}. The $n$-Torus $\mathcal{T}^n$ can be described as the product of $n$ circles $\mathcal{T}^n = \underbrace{\mathcal{S}^1 \times \cdots \times \mathcal{S}^1}_n$.
Each subspace $\mathcal{S}^1$ of the latent space $\mathcal{T}^n$ can be treated independently, so that the posterior distribution can be factorized into a product of $\text{vMF}$ distributions over each latent angle. Tori are a very important class of neural manifolds, e.g. emerging in the toroidal structure of grid cell representations ~\cite{gardner2022toroidal}.
\vspace{-0.3cm}
\paragraph{Parameterization of the Neural Manifold}

Our VAE's latent space is constrained to be the template manifold $\mathcal{Z}$, \textit{i.e.}, it has the topology of the neural manifold $\mathcal{M}$. We train the VAE with a L2 reconstruction loss and the KL-divergence that is suited for the chosen template manifold. The map $\hat{f}$ of its learned decoder gives a continuous parameterization of the neural manifold: the points $z$ of the template manifold $\mathcal{Z}$ parameterize the points $\hat{f}(z)$ on the neural manifold $\mathcal{M}$. Addressing the lack of uniqueness of the parameterization learned will be the focus of Section~\ref{sec:theory}.



\subsection{Geometry: Learning Neural Curvatures}

We leverage the learned deformation $\hat{f}: \mathcal{Z} \rightarrow \mathcal{M} \subset \mathcal{X}$ to extract geometric signatures of the neural manifold $\mathcal{M}$. Importantly, we compute its extrinsic curvature, which represents its \textit{neural shape} in the high-dimensional data space. We rely on Riemannian geometry, which provides tools to quantify curvatures and geometric structures. Additional background on geometry can be found in \cite{aubin1998some}.

\begin{definition}[Riemannian metrics and manifolds~\cite{aubin1998some}]

Let $\mathcal{Z}$ be smooth connected manifold and $T_{z}\mathcal{Z}$ be its tangent space at the point $z \in \mathcal{Z}$. A \textit{Riemannian metric} $g$ on $\mathcal{Z}$ is a collection of positive-definite inner products $g_{z} : T_{z}\mathcal{Z} \times T_{z}\mathcal{Z} \rightarrow \mathbb{R}$ that vary smoothly with $z$. A manifold $\mathcal{Z}$ equipped with a Riemannian metric $g$ is called a Riemannian manifold $(\mathcal{Z}, g)$.
\end{definition}

\begin{figure}
    \centering
    \includegraphics[width=1\linewidth]{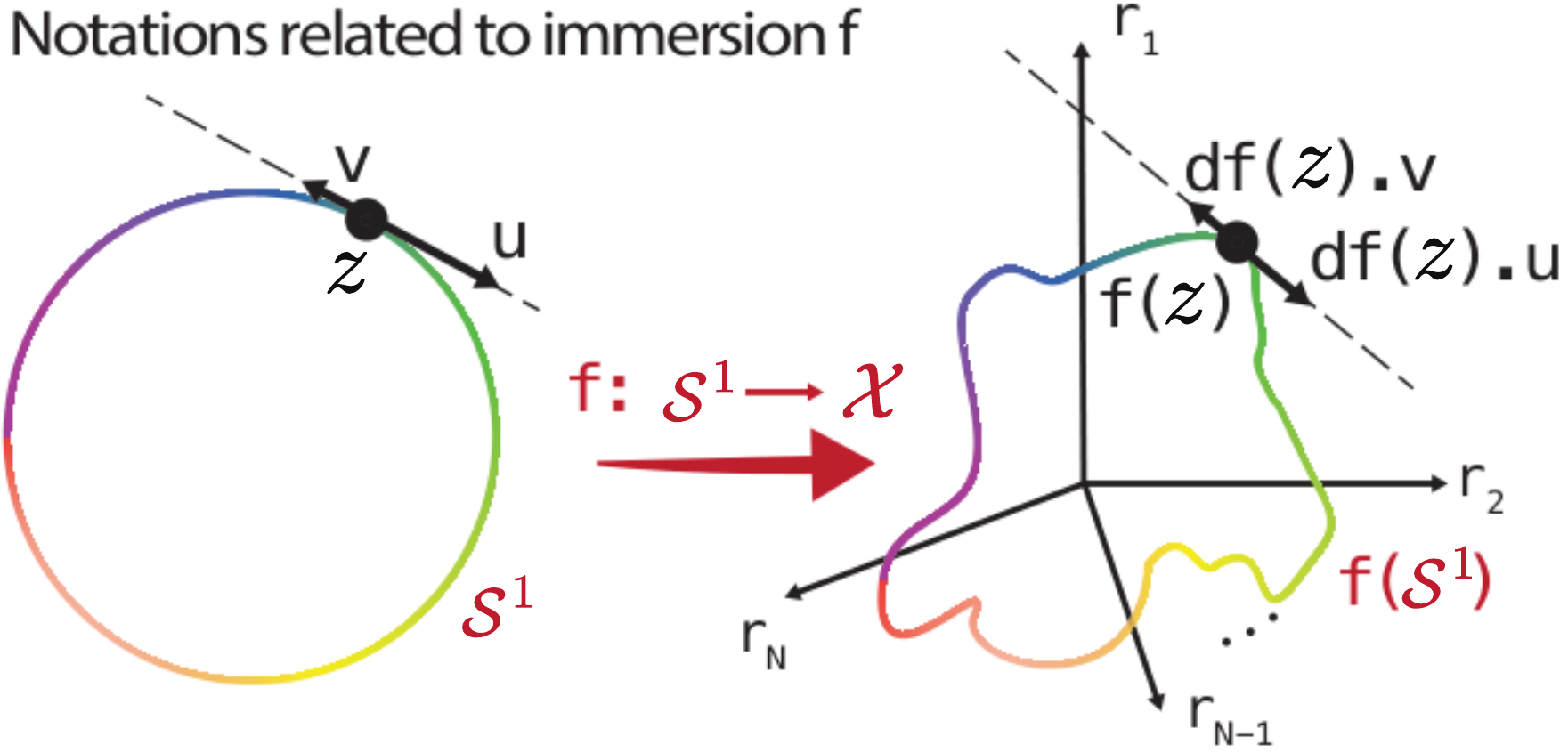}
    \caption{Immersion $f$ sending $\mathcal{Z} = \mathcal{S}^1$ into $f(\mathcal{S}^1)$ immersed in $\mathbb{R}_{+}^N$. The function $f$ maps a point $z \in \mathcal{S}^1$ into a point $f(z) \in f(\mathcal{S}^1)$. Its differential $df(z)$ at $z$ maps vectors $u, v$ tangent to $\mathcal{S}^1$ at $z$, into vectors tangent to $f(\mathcal{S}^1)$ at $f(z)$.}
    \label{fig:pullback}
\end{figure}

We learn the Riemannian metric of a neural manifold $\mathcal{M}$ to learn its geometry. To do so, we represent the neural manifold as a \textit{parameterized} high-dimensional surface in the neural state space $ \mathcal{X} = \mathbb{R}_{+}^N$: a representation given by the learned deformation $\hat{f}$, which is an immersion (see Fig.~\ref{fig:pullback}).

\begin{definition}[Immersion ~\cite{aubin1998some}]\label{def:immersion}
Let $f$ be a smooth map $f: \mathcal{Z} \to \mathcal{X}$ between smooth manifolds $\mathcal{Z}$ and $\mathcal{X}$. The map $f$ is an immersion if its differential at $z$, i.e. the map $ df_{z}: T_{z}\mathcal{Z} \to T_{f(z)}\mathcal{X}$, is injective $\forall z \in \mathcal{Z}$.
\end{definition}

To satisfy the injectivity requirement, we can see that $f$ is an immersion only if $\text{dim}(\mathcal{Z}) \leq N$. A trivial example of an immersion is the injection $(x_{1},..., x_{d}) \mapsto 
(x_{1},...,x_{d},0,...,0)$ from $\mathbb{R}^d$ to another vector space $\mathbb{R}^{N>d}$. By using a decoder $\hat{f}$ that is twice differentiable, via neural network activation functions such as tanh$(\cdot)$ and softplus$(\cdot)$, the learned map $\hat{f}$ is an immersion. Figure~\ref{fig:pullback} shows an immersion $f$ going from the template manifold $\mathcal{Z}=\mathcal{S}^1$ to the neural manifold $\mathcal{M} \subset \mathcal{X}$. The map $f$ parameterizes the manifold $f(\mathcal{S}^1)$ immersed in $\mathcal{X}$ with angular coordinates in $\mathcal{S}^1$.

\vspace{-0.3cm}
\paragraph{Pullback Riemannian Metric} The immersion $\hat{f}$ induces a Riemannian metric structure on $\mathcal{M}$, or equivalently on its template manifold $\mathcal{Z}$.

\begin{definition}[Pullback metric ~\cite{aubin1998some}]\label{def:pullback}
Let $f$ be an immersion $f: \mathcal{Z} \to \mathcal{X}$ where $\mathcal{X} = \mathbb{R}_{+}^N$ is equipped with a Riemannian metric $g$. The pullback metric $f^{*}g$ is a Riemannian metric on $\mathcal{Z}$ defined $\forall z \in \mathcal{Z}$ and $\forall, u, v \in T_{z}\mathcal{Z}$ as:
\begin{align*}
    (f^{*}g)_{z}(u, v)
    &= g_{f(z)}(df(z).u,df(z).v) 
\end{align*}
\end{definition}


Intuitively, the pullback metric quantifies what the neural manifold ``looks like'' in $\mathcal{X} = \mathbb{R}_{+}^N$, i.e. its neural shape in neural state space. In this paper, we choose $g$ to be the canonical Euclidean metric of $\mathcal{X}$: in other words, we quantify the geometry of the neural manifold $\mathcal{M}$ by using (or \textit{pulling back}) the Euclidean metric of $\mathcal{X}$ on $\mathcal{M}$. In practice, the pullback metric is computed with automatic differentiation from $\hat f$. To further provide intuition on definition~\ref{def:pullback}, the derivations of pullback metrics of the common neural manifolds in the first row of Fig.~\ref{fig:topo-versus-geom} are given in the supplementary materials. 
\vspace{-0.3cm}
\paragraph{Extrinsic Curvature}

The Riemannian metric of the neural manifold allows us to calculate geometric quantities like angles, distances, areas, and various types of curvatures. Riemannian geometry applied to deep learning has focused on intrinsic notions of curvatures, given by the Riemann curvature tensor and contractions thereof, such as the Ricci tensor or the scalar curvature tensor ~\cite{hauser2017principles,shao2018riemannian,kuhnel2018latent}. However, we argue that extrinsic curvatures contain more meaningful information, specifically in the context of neural manifolds. Indeed, intrinsic curvatures cannot provide an interesting description of the local geometry of one-dimensional neural manifolds such as rings, as their value is identically zero at each point of such neural manifold (see proof in the appendices). Since one-dimensional neural manifolds are ubiquitous in neuroscience, we suggest instead to use an extrinsic notion of curvature.
 We use one particular extrinsic curvature: the \textit{mean curvature vector}.  We consider a local coordinate system on $\mathcal{Z}$ around $z$ and index its basis elements with $i, j, k$. We will use a local coordinate system on $\mathcal{M} = f(\mathcal{Z}) \subset \mathcal{X}$ around $f(z)$ and index its basis elements with $\alpha, \beta, \gamma$. These notations are used in Definition~\ref{def:h}, and in the appendices, together with Einstein summation convention for repeated indices.

\begin{definition}[Mean curvature vector $H$]\label{def:h}
Consider the immersion $f$ of Definition~\ref{def:immersion} representing the manifold $\mathcal{M} \subset \mathcal{X} = \mathbb{R}_{+}^N$ parameterized by $\mathcal{Z}$. The mean curvature vector $H(z) \in \mathbb{R}^N$ is defined, for $\alpha \in \{1, ..., N\}$, as:
\begin{align*}
    H^{\alpha}(z) 
    &= \frac{1}{N}(f^{*}g)_{z}^{ij}(\partial_{ij}^{2}f^{\alpha}(z) - \Gamma_{ij}^{k}(z)\partial_{k}f^{\alpha}(z)),
\end{align*}
where $(f^{*}g)_{z}^{ij}$ is the inverse of the Riemannian (pullback) metric matrix of Definition~\ref{def:pullback}, $\Gamma_{ij}^k$ are the Christoffel symbols associated with the metric $f^{*}g$.
\end{definition}

This definition is only valid upon choosing $g$ to be the canonical Euclidean metric of $\mathbb{R}_{+}^N$. The general formula for any metric $g$ is in the supplementary materials together with the derivation of Definition~\ref{def:h} in the Euclidean case, and of the mean curvature vectors in the special cases of two-dimensional surfaces in $\mathbb{R}^3$. 


Intuitively, the mean curvature vector at a given point of the manifold $z$ is orthogonal to the manifold with a norm inversely proportional to the radius of the best fitting sphere at that point $z$: spheres of small radii have high curvatures and vice-versa. The mean curvature vector leads us to define the quantity that we use to visually represent the geometric structure of neural manifolds from our experiments: the \textit{curvature profile}.

\begin{definition}[Curvature profile]\label{def:profile}
Consider the immersion $f$ of Definition~\ref{def:immersion} representing the manifold $\mathcal{M} \subset \mathcal{X}$ parameterized by $\mathcal{Z}$. Consider a latent variable $z_0 \in \mathcal{Z}$ and a unit tangent vector $u_0 \in T_{z_0}\mathcal{Z}$. We define the curvature profile of $\mathcal{M}$ in direction $df(z_0).u_0$ as the map from $\mathbb{R}$ to $\mathbb{R}$:
    \begin{align*}
        s \rightarrow \|H(z(s))\|_{z(s)} \quad \text{for: } z(s)= \text{Exp}^f_{z_0}(s.u_0),
    \end{align*}
where $\text{Exp}^f_{z_0}$ is the Riemannian exponential on $\mathcal{Z}$ corresponding to the pullback metric $f^*g$.
\end{definition}

Intuitively, this definition gives the curvature profile of the manifold $\mathcal{Z}$ in any direction $u_0$, as if the manifold was ``sliced'' according to $u_0$. Importantly, the curvature profiles are parameterized with the length $s$ along the geodesic leaving from $z_0$ in the direction of $u_0$. To further provide intuition on Definition~\ref{def:profile}, we give the curvature profiles of the common neural manifolds shown on the first row of Fig.~\ref{fig:topo-versus-geom} in the supplementary materials. In practice, we extend the software Geomstats ~\cite{miolane2020geomstats} by integrating the computation of the mean curvature vector, together with the differential structures of the so-called first and second fundamental forms that provide a general, open-source and unit-tested implementation.

\vspace{-0.1cm}
\section{Theoretical Analysis}\label{sec:theory}

Our method, described in Section~\ref{sec:methods}, yields the \textit{curvature profile} of a neural manifold. If the neural manifold has the geometry of a ``perfect'' circle, sphere, or torus (as in Fig.~\ref{fig:topo-versus-geom}, top row), we expect to find curvature profiles corresponding to the ones computed in the supplementary materials. By contrast, if the neural manifold presents local variations in curvature (as in Fig.~\ref{fig:topo-versus-geom}, bottom row), experiments will give curvature profiles with a different form.

To yield meaningful results for real-world data, the model must additionally meet two theoretical desiderata: invariance under latent reparameterization, and invariance to neuron permutation. Here, we demonstrate that these desiderata are met.


\subsection{Invariance under Latent Reparameterizations}

We demonstrate that our approach does not suffer from the VAE non-identifiability problem of variational autoencoders~\cite{hauberg2019bayes}. By utilizing the geodesic distance $s$ on the latent space in Definition~\ref{def:profile}, we avoid relying on the specific coordinates of the latent variables, resulting in meaningful curvature profiles. 
\vspace{-0.3cm}
\paragraph{VAE Non-Identifiability}\label{sec:VAEnon}

Consider a VAE trained on data $x_i \in \mathcal{M} \subset \mathcal{X}$, which learns the low-dimensional latent variables $z_i \in \mathcal{Z}$ with prior $p(z)$ and the decoder $f: \mathcal{Z} \to \mathcal{X}$ . Consider a reparameterization
\begin{align*}
    \varphi: \mathcal{Z} \to \mathcal{Z}, \qquad 
            z \mapsto \tilde{z} = \varphi(z)  
\end{align*}
with the property that $z \sim p(z)$, $\tilde{z} \sim p(\tilde{z})$, $p(\tilde{z}) = p(z) $. 
Then a VAE with latent variables $\tilde{z}$ and decoder $\tilde{f} = f \circ \varphi^{-1}$ will yield the same reconstruction, and thus be equally optimal. Thus, parameterizing the curvature profile of the neural manifold with a latent variable $z$ will not be meaningful, in the sense that the overall profile will exhibit features that depend on the parameterization $\varphi$ of $\mathcal{Z}$.
This is known as VAE non-identifiability, and it requires us take caution when analyzing the geometry of the latent space~\cite{hauberg2019bayes}.

\paragraph{Invariant Curvature Profile}

We show that the curvature profile defined in Definition~\ref{def:profile} is invariant with respect to reparameterizations $\varphi$.

\begin{lemma}[Invariance under reparameterizations]\label{lem:invariance-reparameterizations}
The curvature profile is invariant under reparameterizations $f \rightarrow f \circ \varphi^{-1}$ of the neural manifold $\mathcal{M}$.
\end{lemma}

We provide a proof in the supplementary materials. The consequence is that our proposed method does not suffer from VAE non-identifiability. An alternative approach to deal with VAE non-identifiability is to directly link each latent variable $z$ with a real-world variable corresponding to its data point $x$. The term $\mathcal{L}_{latent}$ corresponds to the squared distance between the latent variable $z$ and its ground-truth value $z_{gt}$, computed using the Euclidean metric of the embedding space of $\mathcal{Z}$: $\mathcal{L}^{\mathcal{Z}}_{latent} = \text{dist}(z, z_{gt})^2$
where $\text{dist}$ is the geodesic distance associated with the canonical Riemannian metric of $\mathcal{Z}$ (not the pullback metric). This approach is supervised as we enforce that the latent variables $z$ bear meaning related to the real-world task: the local curvature of the neural manifold can be studied in correlations with real-world variables.
This supervised approach effectively selects a canonical parameterization that parameterizes the neural manifold by real-world task variables.

\subsection{Invariance under Neuron Permutation}

In experimental neuroscience, the order in which neurons appear in a data array may change between recording sessions and has no bearing on the underlying latent structure. Thus, methods for extracting geometric structure from neural data should not depend on neuron ordering. In this regard, we first validate our pre-processing step and previous topological methods that rely on TDA by demonstrating the independence of the topology of the neural manifold from neuron order. We also show that the order in which we record the $N$ neurons to form the neural state space $\mathbb{R}_{+}^N$ does not impact the curvature profiles. 

\begin{lemma}[Invariance under permutations]\label{lem:invariance-permutation}
Consider a neural manifold $\mathcal{M}$ embedded in neural state space $\mathbb{R}_{+}^N$ corresponding to the recording of $N$ neurons. Permuting the order of the $N$ neurons: (i) leaves the topology of $\mathcal{M}$ invariant, (ii) leaves the geometry of $\mathcal{M}$ invariant.
\end{lemma}

The proof is in the supplementary materials and relies on the fact that a permutation of the neurons in $\mathbb{R}_{+}^N$ yields an isometry of $\mathcal{M}$, in other words: that the Euclidean metric $g$ of $\mathbb{R}_{+}^N$ is invariant to permutations. 
Consequently, future works aiming at pulling back another metric should first verify that the new metric is invariant to permutations. 


\section{Experiments}\label{sec:experiments}

\subsection{Synthetic Manifolds}

We first test our method on synthetic datasets of circles, spheres and tori distorted by small Gaussian ``bumps'', embedded in $\mathbb{R}_{+}^2$, $\mathbb{R}_{+}^3$ and  $\mathbb{R}_{+}^3$ respectively \textemdash see Fig.~\ref{fig:toy_results}. The exact process generating these datasets in given in the supplementary materials. In these synthetic experiments, the curvature profile can be computed analytically and compared to the learned curvature profile, as in Fig.~\ref{fig:toy_results}. We verify that the ground-truth topology and geometry can be recovered, validate the model's invariance under reparameterization, and test the effects of simulated experimental noise and the number of simulated neurons.
\begin{figure}
    \centering
    \includegraphics[width=1\linewidth]{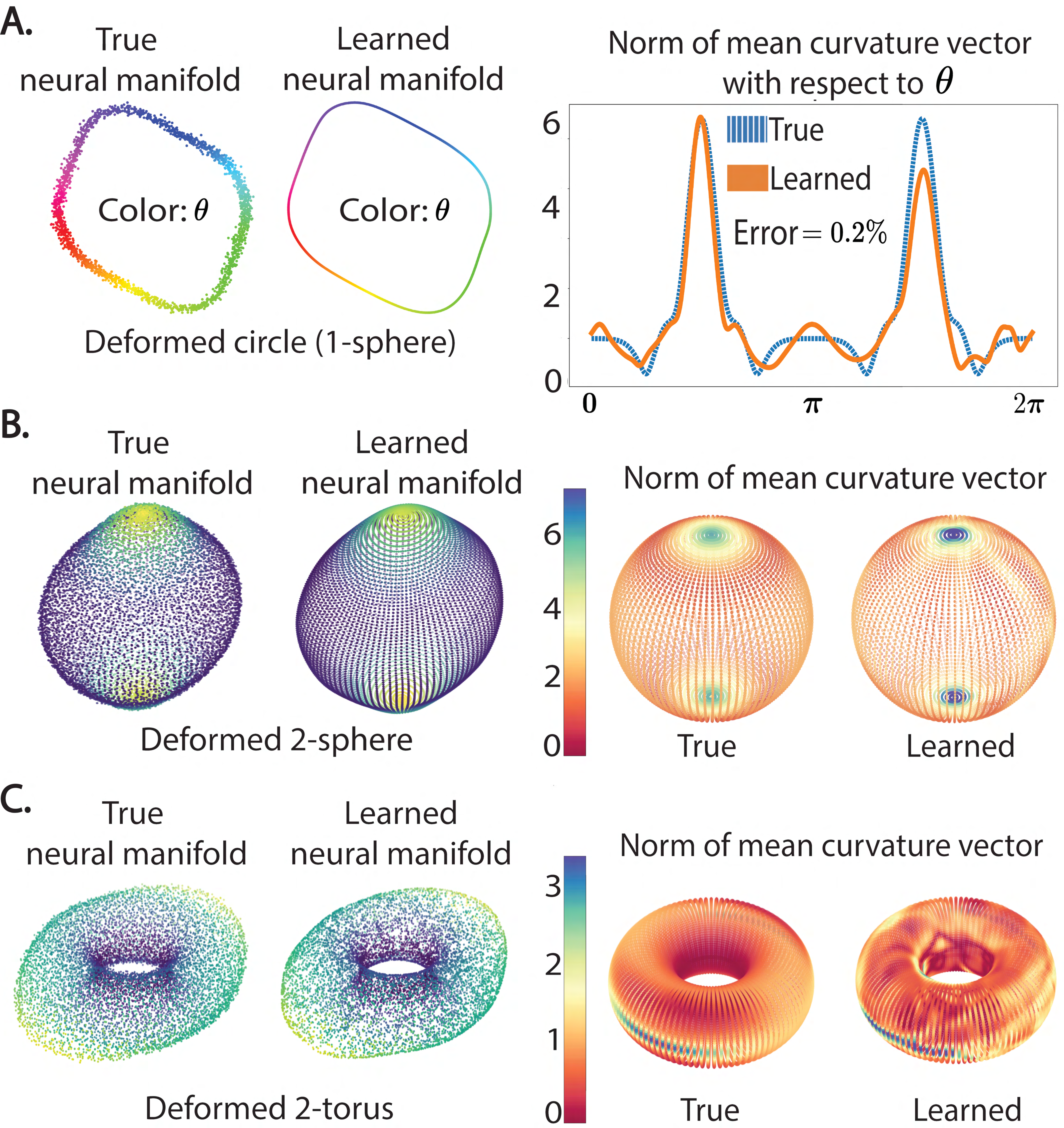}
    \caption{Synthetic manifolds created from smooth deformations of template manifolds $\mathcal{S}^1$, $\mathcal{S}^2$ and $\mathcal{T}^2$. \textbf{A.} Left, synthetic manifold created from deformed circle along with its VAE reconstruction, color represents ground-truth angle. Right, profile of mean curvature norm: ground-truth (blue) compared to estimated (orange) along with the calculated error. \textbf{B} and \textbf{C}. The left side shows the synthetic manifolds and their corresponding reconstructions (color added for visualization of bump-like deformations); the right side shows the norm of the mean curvature plotted on the template manifolds, computed (1) from the ground truth smooth deformation, and (2) from the smooth deformation $\hat f$ learned by the VAE decoder.}
    \label{fig:toy_results}
\end{figure}

\paragraph{Validation of Learned Topology and Geometry} We first verify that TDA correctly learns the topology of the neural manifold in the synthetic datasets. This is illustrated in the supplementary materials. We next verify that our approach correctly learns the geometry of the neural manifolds when supervision is added to enforce a canonical parameterization of the latent space. Our VAE correctly learns a compressed latent representation $z$ in the desired template manifold $\mathcal{Z}$, that further corresponds to the ground-truth $z_{gt}$, as shown in Fig.~\ref{fig:toy_results} A, B for $z = \theta$ representing the angle on $\mathcal{S}^1$ and the latitude on $\mathcal{S}^2$ respectively. The VAE's trained decoder provides an accurate reconstruction of the input neural manifold (see Fig.~\ref{fig:toy_results} (A-B-C) Left).
The trained decoder function then allows us to calculate the local mean curvature at each point on the manifold (see Fig.~\ref{fig:toy_results} (A-B-C) Right).

 \paragraph{Effect of Noise and Number of Neurons} We run experiments on synthetic distorted circles to evaluate the accuracy of the curvature estimation. Our evaluation metric is the curvature estimation error defined as a normalized error proportional to the integrated difference squared of the two mean curvature profiles:

 \begin{definition}[Curvature Estimation Error] The error between the estimated curvature $\hat{H}$ and its true value $H$ is given by:
\begin{equation*}\label{eq:error}
\text{Error}(H, \hat{H}) = \frac{\displaystyle\int_{\mathcal{Z}}
\|H(z)-\hat{H}(z)\|^2 \,d\mathcal{Z}(z)}
{\displaystyle\int_{\mathcal{Z}} (\|H(z)\|^2 + \|\hat{H}(z)\|^2)
\,d\mathcal{Z}(z)},
\end{equation*}
where $\mathcal{Z}$ is the known template manifold and $d\mathcal{Z}(z)$ its Riemannian measure.
 \end{definition}
 

The curvature estimation error is thus expressed as a fraction (\%) of the sum of both curvatures. We compute $\text{Error}(H, \hat{H})$ for curvature profiles estimated from a wide range of distorted circles. Specifically, we generate synthetic distorted circles as outlined in the supplementary materials, where we vary: (i) the noise level, i.e., the standard deviation of the Gaussian noise: $\sigma \in [0\%,  12\%]$ expressed in \% of the radius of the distorted circle, i.e. the manifold's size; and (ii) the number of recorded neurons $N$, which is the embedding dimension of the synthetic manifold: $N \in \left[3, 25\right]$. We emphasize that we chose these noise values in order to match the noise amplitudes observed in real neural data\cite{jayakumar2019recalibration}, and we chose values for $N$ that correspond to the number of neurons traditionally recorded in real experimental neuroscience. 

\begin{figure}
    \centering
    \includegraphics[width=1\linewidth]{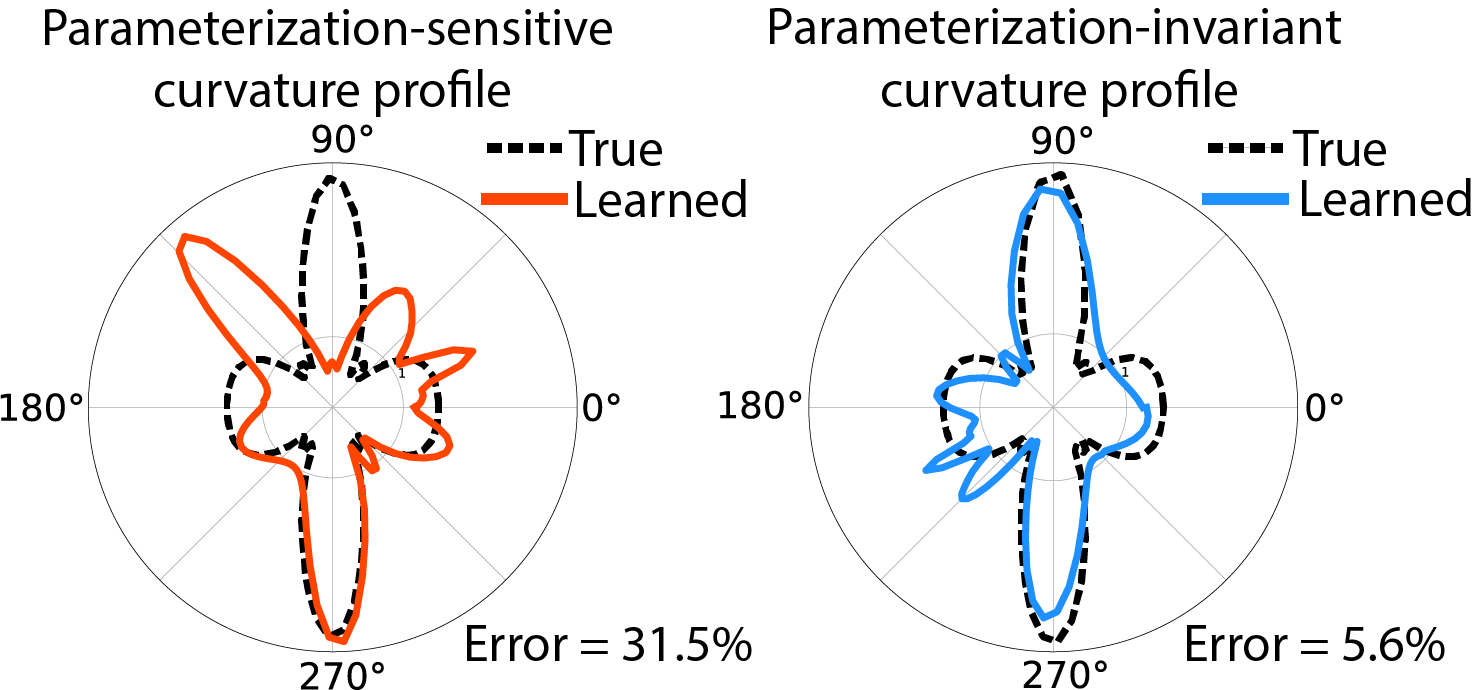}
    \caption{Computed curvature profiles of learned decoder immersion, plotted on the a circle of radius $1/2$. Left: We plot the magnitude of the mean curvature as a function of the latent angle $\theta$, $\norm{H} = \norm{H(\theta)}$. This curvature profile suffers from VAE non-identifiability. Right: We plot the magnitude of the mean curvature as a function of the distance of a latent angle $\theta$ to some reference angle $\theta_0$, $\norm{H} = \norm{H(\text{dist}(\theta_0,\theta))}$, where $\text{dist}$ is computed using the pullback metric. This approach is significantly better able to capture the true curvature structure on the neural manifold $\mathcal{S}^1$}
    \label{fig:s1_repam_inv}
\end{figure}

Fig.~\ref{fig:error_est} (A) in the supplementary materials shows that $\text{Error}(H, \hat{H})$ varies between 0.5\% and 5\% of the true curvature for both distorted circles across these realistic noise levels when the number of neurons is fixed at $N=2$. For each value of $\sigma$, a synthetic manifold was created and our method was applied 5 times: the vertical orange error bars show the minimum and maximum of the error across 5 trainings. Fig.~\ref{fig:error_est} (B) shows that $\text{Error}(H, \hat{H})$ does not depend on the number of neurons $N$, and specifically does not reach the magnitude of error observed upon varying $\sigma$. 
This demonstrates that our method is suited to learn the geometry of neural manifolds across these realistic noise levels and number of neurons, and motivate its use in the next section on simulated and experimental one-dimensional neural manifolds.
\vspace{-2pt}
\paragraph{Validation of Invariance under Reparameterizations} As shown in Fig.~\ref{fig:s1_repam_inv}, we experimentally demonstrate that in the unsupervised setting, our parameterization-invariant approach for computing the curvature profile of distorted circles is better able to recover the true curvature structure on the synthetic manifold, compared to the naive parameterization-sensitive procedure.

\subsection{Neural Manifolds}

\subsubsection{Simulated Place Cells}

\begin{figure}
    \centering
    \vspace{-0.2cm}
    \includegraphics[width=0.85\linewidth]{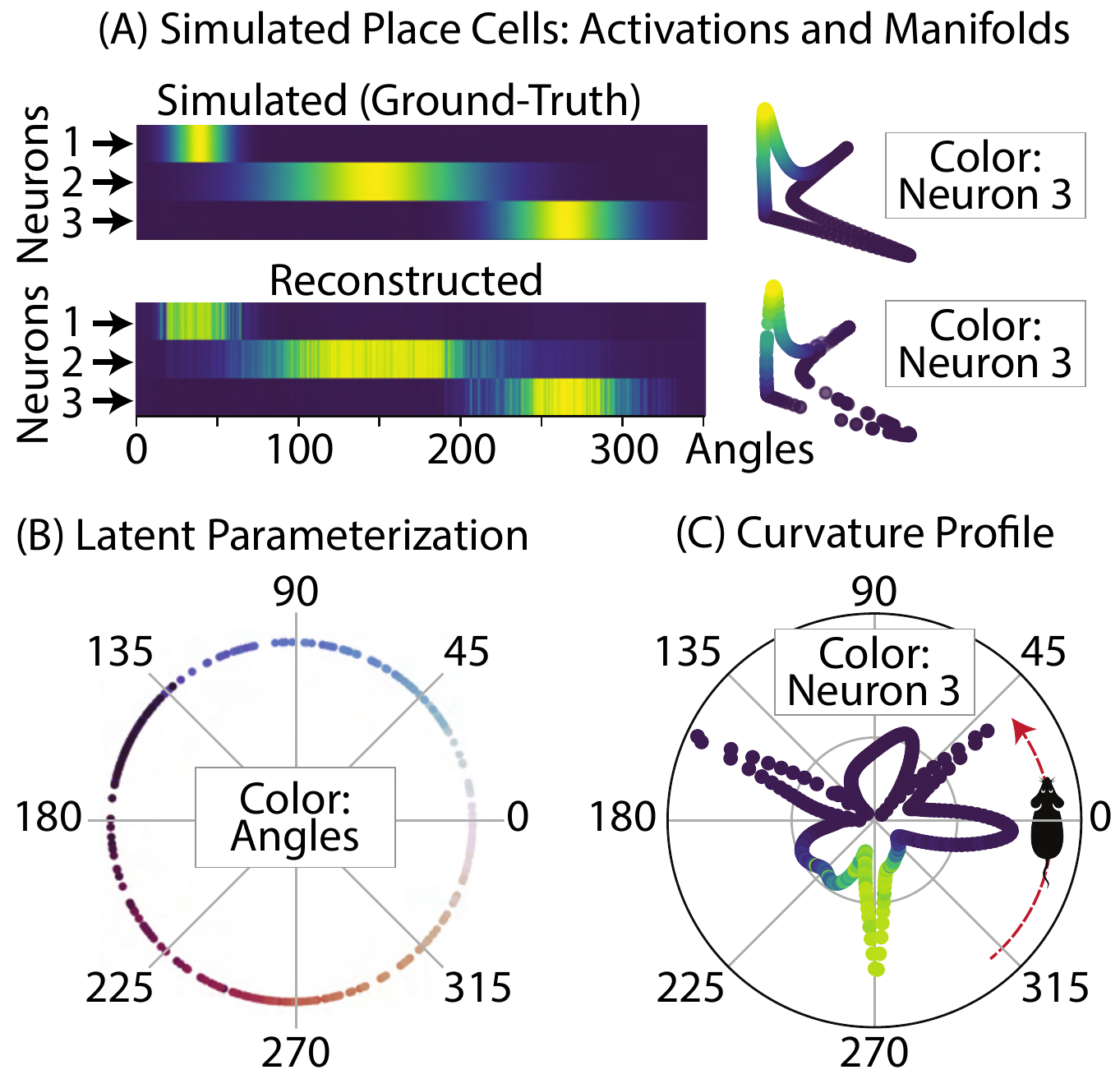}
    \caption{Simulated place cells. (A) Left: Simulated and reconstructed neural activity with respect to the positional angles of the animal in lab space. Right: Simulated and reconstructed neural manifolds, colored with the simulated and reconstructed activations of neuron 3. (B) Latent parameterization: angular latent variables are colored by the animal's positional angles in lab space. (C) Curvature profile in log scale: angles represent the animal's positional angles, colored by the reconstructed activation of neuron 3.}
    \label{fig:three_place_cells}
\end{figure}

We simulate the neural activity of place cells to provide intuition about the neural geometry that we expect to find in experimental data (next section). In this simulation, an animal moves along a circle in lab space. We simulate the activity of 3 place cells as shown in Fig.~\ref{fig:three_place_cells} (A): each neuron peaks when the animal is at a specific lab angle on the circle (40 degrees, 150 and 270 degrees respectively) and has a place field of a fixed width (80, 300, and 180 respectively). These neural recordings form the neural manifold $\mathcal{M} \subset \mathbb{R}_{+}^3$ shown in Fig.~\ref{fig:three_place_cells} (A) (Right - Simulated). In this experiment, we can record the true positional angle of the animal with cameras provided in the lab. Therefore, we use a canonical parameterization of the latent space's angles with latent loss term from the previous section.

Our approach correctly reconstructs the neural activity and learns a neural manifold whose geometry matches the ground-truth in Fig.~\ref{fig:three_place_cells} (A). The canonical parameterization of the latent space is correctly learned thanks to the supervision of the latent loss, see Fig.~\ref{fig:three_place_cells} (B). The curvature profile is shown in Fig.~\ref{fig:three_place_cells} (C) The profile shows 3 shallow peaks, which correspond to the higher curvature observed in the neural manifold when the animal is at at angle that is in between the peaks of two place fields.
\vspace{-2pt}
\subsubsection{Experimental Place Cells}\label{sec:expt_place_cells}

We apply our method to real neural data from rats~\cite{jayakumar2019recalibration} running in a VR Dome apparatus~\cite{madhav2022dome}, which realizes the simulations from the previous subsection. In this experiment, animals run on a circular track $\mathcal{S}^1$ surrounded by projected visual cues, while we record the neural activity of place cells in the hippocampal CA1 region. As place cells typically encode the position of the rat in space, we expect the topology of the neural manifold to be $S^1$ and choose $\mathcal{Z}=\mathcal{S}^1$. This experiment also possesses a ``canonical'' parameterization as the animal's positional angle is recorded by a lab camera: we use this angle to supervise the latent angles with the latent loss.
We discuss the details of the experimental place cell data in the supplementary materials. Here, we detail how our approach can be used to reveal novel neuroscience insights. In the Dome experiment, visual landmarks were moved by a fraction ($G$) of the rat's physical movement on the circular track~\cite{jayakumar2019recalibration}. Place cells remain locked to these moving landmarks - i.e. the radius of the $\mathcal{S}^1$ neural manifold scales according to $G$. This scaling persists even after landmarks are turned off, indicating a recalibration of self-motion inputs. In the original work, the radius of the neural manifold ($\propto$ inverse of curvature $H$) was determined through a Fourier-based method, necessitating multiple cycles through the manifold to generate an averaged estimate. Our method can track the sub-cycle evolution of $H$, allowing for more precise understanding of how local sensory inputs (such as proximal landmarks) contribute to the dynamical scaling of neural geometry. A subsequent experiment showed how the latent position encoded in place cells can be decoupled from physical space in the absence of visual landmarks~\cite{madhav2022closed-loop}. Our method can be used to decode the local curvature profile of the neural representation in the absence of correlated real-world variables. Recent research shows that neurons of the cognitive map can encode task-relevant non-spatial variables~\cite{aronov2017mapping, nieh2021geometry, knudsen2021hippocampal, danjo2018spatial, omer2018social}. Our method can be used to test whether the geometric features of these latent variables correspond to that of the task.

\begin{figure}
    \centering
    \vspace{-0.2cm}
    \includegraphics[width=0.9\linewidth]{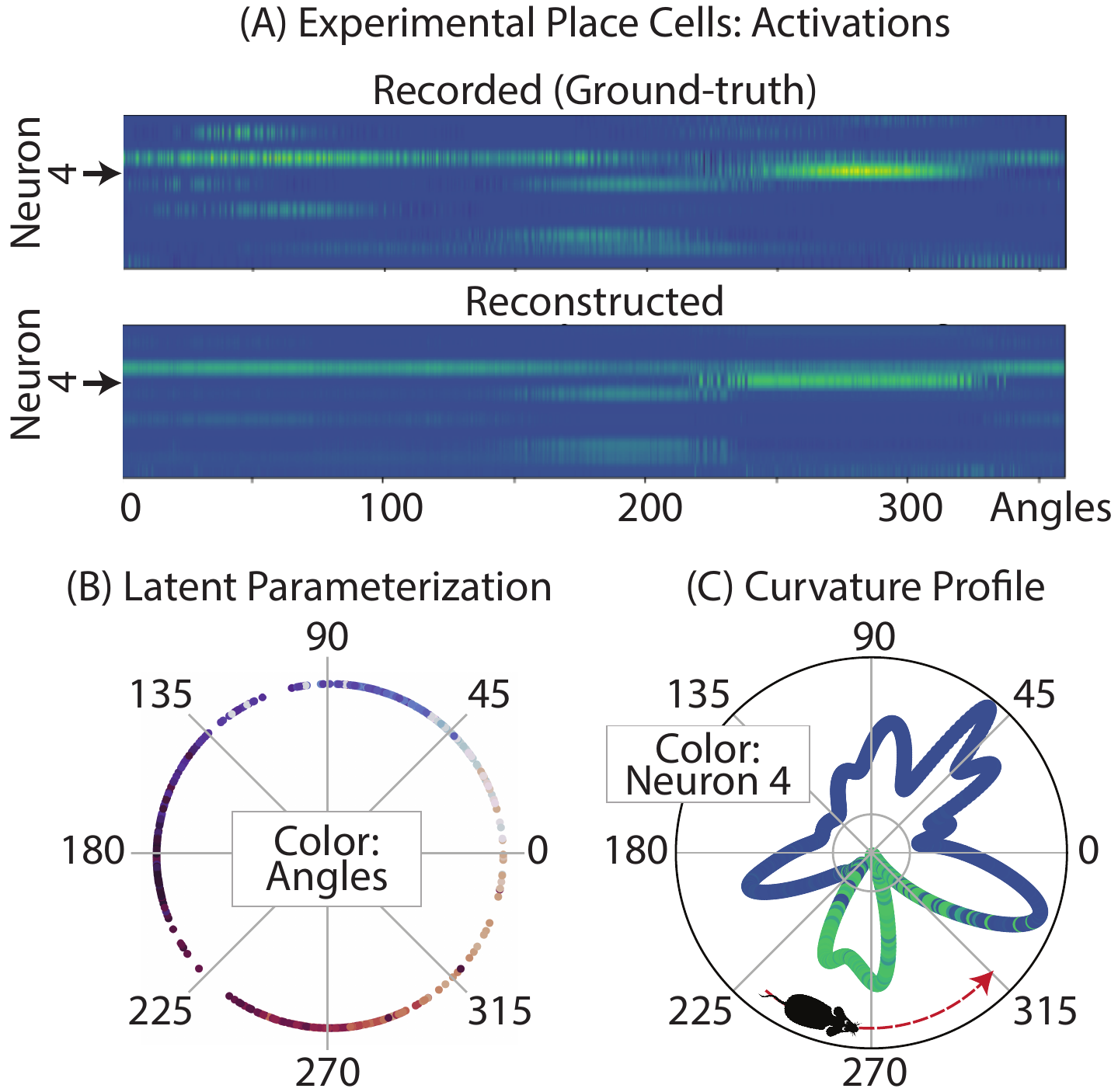}
    \caption{Experimental Place Cells. (A) Recorded versus reconstructed neural activity with respect to the animal's positional angles in lab space. (B) Latent parameterization: angular latent variables are colored by animal's positional angles in lab space. (C) Curvature profile in log scale: angles represent animal's positional angles colored by the reconstructed activation of neuron 4.}
    \label{fig:expt41}
\end{figure}

\section{Conclusion}
We have introduced a novel approach for quantifying the geometry of neural manifolds. We expect that this method will open new avenues for research in the geometry of neural representations.

\section{Acknowledgements}
We would like to thank James Knierim, Noah Cowan and Ravikrishnan Jayakumar at Johns Hopkins University for providing experimental data. The rodent place cell data analyzed in this manuscript was collected in the Knierim and Cowan labs by R. Jayakumar and M. Madhav. We also thank David Klindt for insightful discussions and comments on the manuscript. 

{\small
\bibliographystyle{ieee_fullname}
\bibliography{references}
}

\appendix
\input{supp.tex}

\end{document}

%% file: supp.tex
\section{Variational Autoencoders}

This section reviews VAEs and the tools of Riemannian geometry that support our curvature estimation method. For further background in variational inference and VAEs, we direct the reader to \cite{doersch2021tutorial,blei2017variational}.

A \textit{variational autoencoder} (VAE) is a generative deep latent variable model widely used for unsupervised learning ~\cite{kingma2013auto}. A VAE uses an autoencoder architecture to implement variational inference: it is trained to encode input data in a compact latent representation, and then decode the latent representation to reconstruct the original input data. Consider an $N$-dimensional dataset of $k$ vectors $x_1, \ldots,  x_k \in \mathbb{R}^N$. The VAE models each data vector $x_i$ as being sampled from a likelihood distribution $p(x_i|z_i)$ with lower-dimensional unobserved latent variable $z_i$. The likelihood distribution is usually taken to be Gaussian, so we write the reconstructed input as $x^{rec}_i = f(z_i) + \epsilon_i$ with $\epsilon_i \sim \mathcal{N}(0,\sigma^2 \mathbb{I}_N)$. The function $f$ is here represented by a neural network called the \textit{decoder}. 

The VAE simultaneously trains an \textit{encoder} that represents the approximate posterior distribution $q(z|x)$ over the latent variables $z$. 
The VAE achieves its objective by minimizing an upper-bound of the negative log-likelihood, which writes as the sum of a reconstruction loss and a Kullback-Leibler (KL) divergence:
\begin{equation}
\begin{split}
    \mathcal{L} & = \mathcal{L}_{rec} + \mathcal{L}_{KL} \\
    & = - \mathbb{E}_{q(z)}[\log{p(x|z)}] + \infdiv{q(z|x)}{p(z)}.
\end{split}
\end{equation}

We use a similar loss in our experiments, but we adapt the KL term to the topology of the latent space, which we call the template manifold $\mathcal{Z}$.

\section{Derivations of the Second Fundamental Form for Surfaces in 3-D}

We give additional details on the second fundamental form $\mathbb{\RN{2}}$, for a surface within $\mathbb{R}^3$, and the associated mean curvature vector.

\begin{definition}[Second fundamental form $\mathbb{\RN{2}}$ of a surface in $\mathbb{R}^3$]

Consider a surface $S$ in $\mathbb{R}^3$, as the graph of a twice differentiable function $f(u,v)$. We choose our coordinate system such that $f(u,v) = 0$ defines the tangent plane to $S$ at the point $z$ and $(u,v)$ are the coordinates of a small displacement $x \in T_{z}S$. By Taylor's theorem, the best approximation of $f(u,v)$ in a small region around $z$ is
\begin{align*} 
    &f_{z}(u,v) = 
    \frac{1}{2}(L u^{2} + 2M uv + N v^{2}) 
    = \frac{1}{2} x^{T}\mathbb{\RN{2}}_{z} \ x,\\
   &\text{with: } \mathbb{\RN{2}}_{z} = \begin{bmatrix}
L & M \\
M & N 
\end{bmatrix}.
\end{align*}
In this equation, $\mathbb{\RN{2}}_{z}$ is a matrix of second partial derivatives of $f$ at $z$, called the second fundamental form of the surface.
\end{definition}

The second fundamental form allows to define the mean curvature vector.

\begin{definition}[Mean curvature $H$ of a surface in $\mathbb{R}^3$ from its second fundamental form $\mathbb{\RN{2}}$]\label{def:h3second}
Consider a surface in $\mathbb{R}^3$ represented by its second fundamental form $\mathbb{\RN{2}}_{z}$. Then, its mean curvature vector is given by:
\begin{equation}
    H_{z} = \frac{1}{2}\text{Tr} \ \mathbb{\RN{2}}_{z}.
\end{equation}
\end{definition}

In the specific case of two-dimensional surfaces immersed in $\mathbb{R}^3$, the mean curvature vector also enjoys an equivalent definition, given below.

\begin{definition}[Mean curvature $H$ of a surface in $\mathbb{R}^3$]\label{def:h3}
Consider a 2-dimensional surface $S$ embedded in $\mathbb{R}^3$. A normal vector at a point $z \in S$ defines a family of normal planes containing this vector, each of which cuts the surface in a direction $\psi$ producing a plane curve. The curvature of such a curve at $z$ is given by $\kappa_{z} = \frac{1}{R(z)}$, where $R(z)$ is the radius of the osculating circle at that point, i.e. the circle that best approximates this curve locally.

The mean curvature $H_{z}$ at the point $z$ is defined as
\begin{equation}
    H_{z} = \frac{1}{2\pi} \int_{0}^{2\pi} \kappa_{z} (\psi) \, d\psi,
\end{equation}
which is the average of the curvatures $\kappa_{z}(\psi)$ over all directions $\psi$ in the tangent plane at $z$. 
\end{definition}

Definition~\ref{def:h3} provides the intuition behind the name ``mean'' curvature, as its defining equation is effectively a mean.

\section{Derivations of the Mean Curvature Vectors}

\subsection{General Formula}

We present the general definition of mean curvature, that builds on the definition of second fundamental form. We refer the reader to the next subsections for concrete examples of these definitions in the special case of two-dimensional surfaces in $\mathbb{R}^3$.

\begin{definition}[Second fundamental form \cite{aubin1998some}]\label{def:secfun}
Consider the manifold $\mathcal{M}$ represented as the immersion of $\mathcal{Z}$ into $\mathcal{X}$ such that $\mathcal{M} = f(\mathcal{Z})$, $\mathcal{M} \subset \mathcal{X}$. We have:
\begin{align*}
   \mathbb{\RN{2}}(z)_{ij}^\alpha = \nabla^2_{ij}f^\alpha(z) =\partial_{i j}^2 f^\alpha(z)-\Gamma_{i j}^k(z) \partial_k f^\alpha(z)\\
   \qquad\qquad +\bar{\Gamma}_{\beta \gamma}^\alpha(f(z)) \partial_i f^\beta(z) \partial_j f^\gamma(z).
\end{align*}
Here, $\Gamma_{i j}^k$ are the Christoffel symbols of $\mathcal{Z}$ for the pullback metric and $\bar{\Gamma}_{\beta \gamma}^\alpha$ the Christoffel symbols of $\mathcal{X}$ for the metric of $\mathcal{X}$. In this formula, $i, j$ are indices for basis elements of $T_{f(z)}\mathcal{X}$, identified with basis elements of $T_z \mathcal{Z}$ since both tangent spaces share the same metric; while $\alpha$ in an index for a basis element of $N_{f(z)}\mathcal{X}$. 
\end{definition}

We note that, in the case where $\mathcal{X} = \mathbb{R}^N$ or $\mathcal{X} = \mathbb{R}_{+}^N$ , the Christoffel symbols $\bar{\Gamma}_{\beta \gamma}^\alpha$ are all zeros. Additionally, in the specific case where the manifold $\mathcal{Z}$ is one dimensional, its Christoffel symbols are 0. In other words, for a ring immersed in $\mathbb{R}_{+}^N$, the Hessian with respect to the pullback metric is the traditional Hessian: $\nabla^2_{ij}f(z)= \frac{\partial^2 f}{\partial x_i \partial x_j}(z)$. 

We now give the general definition of the mean curvature vector, for any submanifold of $\mathcal{X}$.

\begin{definition}[Mean curvature vector \cite{aubin1998some}]
The mean curvature vector $H(z)$ of $\mathcal{M} = f(\mathcal{Z}) \subset \mathcal{X}$ is defined as:
\begin{align*}
H^\alpha(z)
    = \frac{1}{N} \text{Tr} \ \mathbb{\RN{2}}(z)^\alpha 
    = \frac{1}{N} g^{ij}{\mathbb{\RN{2}}(z)}_{ij}^\alpha,
\end{align*}
where $N$ is the dimension of $\mathcal{X}$, and the trace $Tr$ is computed with respect to $g^{ij}$, the inverse of the Riemannian metric matrix of $\mathcal{Z}$.
\end{definition}

This leads us to the definition of mean curvature vector of an immersed manifold.

\begin{definition}[Mean curvature vector (immersed manifold) \cite{aubin1998some}]
The mean curvature vector $H(z)$ of $\mathcal{M}$ is defined as:
\begin{align*}
H^\alpha(z) 
    &= \frac{1}{N} g^{ij}\Big(\partial_{i j}^2 f^\alpha(z)-\Gamma_{i j}^k(z) \partial_k f^\alpha(z)  \\
    &\qquad\qquad  +\bar{\Gamma}_{\beta \gamma}^\alpha(f(z)) \partial_i f^\beta(z) \partial_j f^\gamma(z)\Big),
\end{align*}
where $N$ is the dimension of $\mathcal{X}$, and $g^{ij}$ is the inverse of the Riemannian metric matrix of $\mathcal{Z}$.
\end{definition}

\subsection{Mean Curvatures of the Circle} %

\begin{example}[Mean curvatures of the circle immersed in $\mathbb{R}^N$]
We consider a circle $C$ of radius $R$ immersed in $\mathbb{R}^N$. The norms of its mean curvature vector is:
\begin{equation}
    \|H_C(\theta)\| = \frac{1}{R}, \quad \forall \theta \in \mathcal{S}^1.
\end{equation}
\end{example}

\begin{proof}
We compute the mean curvature of a circle immersed in $\mathbb{R}^N$ as:
\begin{align*}
   f: \mathcal{S}^1 &\mapsto \mathbb{R}^N \\
   \theta &\mapsto f(\theta) = P. 
   \begin{bmatrix}
   R\cos \theta \\
   R\sin \theta \\
   0 \\
   \vdots \\
   0
   \end{bmatrix} + t,
\end{align*}
where $P \in SO(N)$ represents a rotation in $\mathbb{R}^N$, and $t \in \mathbb{R}^N$ a translation in $\mathbb{R}^N$ illustrating that the circle can be placed and oriented in any direction in $\mathbb{R}^N$. 

We compute the second fundamental form, for $k =1, 2, 3$:
\begin{equation}
    \mathbb{\RN{2}}_{11}(\theta) 
    = 
    \begin{bmatrix}
    \frac{d^2f^1}{d\theta^2 }(\theta) \\
    \frac{d^2f^2}{d\theta^2 }(\theta) \\
    \vdots \\
    \frac{d^2f^N}{d\theta^2 }(\theta)
    \end{bmatrix}
    = \begin{bmatrix}
    -R \cos \theta \\
    - R \sin \theta \\
    0 \\
    \vdots \\
    0
    \end{bmatrix}.
\end{equation}

The mean curvature vector is then:
\begin{align*} 
    H_C(\theta) 
    &= \frac{1}{1}\text{Tr}
        \mathbb{\RN{2}}_{\theta}\\
    &= g^{11}
    {\mathbb{\RN{2}}_{11}}(\theta)\\
    &= \frac{1}{R^2}
    \begin{bmatrix}
    -R \cos \theta \\
    - R \sin \theta \\
    0 \\
    \vdots \\
    0
    \end{bmatrix}\\
    &= \frac{1}{R} 
    \begin{bmatrix}
    - \cos \theta \\
    - \sin \theta \\
    0 \\
    \vdots \\
    0
    \end{bmatrix}
\end{align*}
Its norm is: $\|H_C(\theta)\| = \frac{1}{R}$ for all $\theta \in \mathcal{S}^1$.
\end{proof}

\subsection{Mean Curvatures of the Sphere}
\begin{example}[Mean curvatures of the sphere immersed in $\mathbb{R}^N$]
We consider a sphere $S$ of radius $R$ immersed in $\mathbb{R}^N$. The norm of its mean curvature vector is: 
\begin{equation}
   \|H_C(\theta, \phi)\| = \frac{1}{R}, \qquad \forall \theta, \phi \in \mathcal{S}^2.
\end{equation}
\end{example}

\begin{proof}
We compute the mean curvature of a sphere of radius $R$ immersed in $\mathbb{R}^N$ as:
\begin{align*}
   f: \mathcal{S}^2 &\mapsto \mathbb{R}^N \\
   \theta, \phi &\mapsto f(\theta, \phi) = P. 
   \begin{bmatrix}
    R\sin \theta \cos \phi \\
    R\sin \theta \sin \phi \\
    R \cos \theta \\
    0 \\
    \vdots \\
    0
   \end{bmatrix} + t,
\end{align*}
where $P \in SO(N)$ represents a rotation in $\mathbb{R}^N$, and $t \in \mathbb{R}^N$ a translation in $\mathbb{R}^N$ illustrating that the sphere can be placed and oriented in any direction in $\mathbb{R}^N$. 

We compute the Hessian:
\begin{equation}
   \frac{\partial^2 f}{\partial x_i \partial x_j}(\theta, \phi)  
    = \begin{bmatrix}
     \frac{\partial^2f^1}{\partial x_i \partial x_j }(\theta, \phi) \\
     \frac{\partial^2f^2}{\partial x_i \partial x_j }(\theta, \phi) \\
     \frac{\partial^2f^3}{\partial x_i \partial x_j }(\theta, \phi) \\
     0 \\
     \vdots \\
     0
    \end{bmatrix},
\end{equation}
where we use the conventions $x_1 = \theta$ and $x_2 = \phi$. In what follows, for conciseness of the derivations, we do not write the components $\alpha$ of $f^\alpha$ for $\alpha = 4, ..., N$, as they only contribute terms equal to 0.

We get:
\begin{align*}
\frac{\partial^2 f}{\partial \theta^2}(\theta, \phi) &= \begin{bmatrix}
     - R \sin \theta \cos\phi \\
     - R \sin \theta \sin \phi \\
     - R \cos \theta
\end{bmatrix},\\
\frac{\partial^2 f}{\partial \theta \partial \phi}(\theta, \phi) &= \begin{bmatrix}
     - R \cos \theta \sin\phi \\
     R \cos \theta \cos \phi \\
     0 \\
     0 \\
     \vdots \\
     0
\end{bmatrix},\\
\frac{\partial^2 f}{\partial \phi^2}(\theta, \phi) &= \begin{bmatrix}
     - R \sin \theta \cos\phi \\
     - R \sin \theta \sin\phi \\
     0 
\end{bmatrix}.
\end{align*}

We only compute the diagonal terms, avoiding the computation of $\frac{\partial^2 f}{\partial \theta \partial \phi}(\theta, \phi)$ because we only need the diagonal terms in the definition of the trace, given that the inverse of the pullback metric is diagonal.

We compute the Hessian with respect to the pullback metric, again omitting its components for $\alpha > 3$.
\begin{equation}
    \mathbb{\RN{2}}_{ij}(\theta, \phi) 
    = \begin{bmatrix}
     \frac{\partial^2f^1}{\partial x_i \partial x_j }(\theta, \phi) - \sum_{k=1}^2 \Gamma_{ij}^k \frac{\partial f^1}{\partial x_k} \\
     \frac{\partial^2f^2}{\partial x_i \partial x_j }(\theta, \phi) - \sum_{k=1}^2 \Gamma_{ij}^k \frac{\partial f^2}{\partial x_k}  \\
     \frac{\partial^2f^3}{\partial x_i \partial x_j }(\theta, \phi) - \sum_{k=1}^2 \Gamma_{ij}^k \frac{\partial f^3}{\partial x_k}
    \end{bmatrix}.
\end{equation}

For the 2-sphere, the Christoffel symbols are:
\begin{align*}&\Gamma_{11}^1=\Gamma_{11}^2=\Gamma_{22}^2=\Gamma_{12}^1=\Gamma_{21}^1=0,\\
&\Gamma_{22}^1=-\sin \theta \cos \theta,\\
&\Gamma_{12}^2=\Gamma_{21}^2=\frac{\cos \theta}{\sin \theta},
\end{align*}

so that we get $ \mathbb{\RN{2}}_{11}(\theta, \phi) $:
\begin{align*}
 \mathbb{\RN{2}}_{11}(\theta, \phi) 
 &= \begin{bmatrix}
     - R \sin \theta \cos\phi \\
     - R \sin \theta \sin \phi \\
     - R \cos \theta
\end{bmatrix} -  \Gamma_{11}^1 \frac{\partial f}{\partial x_1} - \Gamma_{11}^2 \frac{\partial f}{\partial x_2}  \\
  &= \begin{bmatrix}
     - R \sin \theta \cos\phi \\
     - R \sin \theta \sin \phi \\
     - R \cos \theta
\end{bmatrix} -  0. \frac{\partial f}{\partial x_1} - 0. \frac{\partial f}{\partial x_2}  \\
  &= \begin{bmatrix}
     - R \sin \theta \cos\phi \\
     - R \sin \theta \sin \phi \\
     - R \cos \theta
\end{bmatrix},
\end{align*}
as well as $ \mathbb{\RN{2}}_{22}(\theta, \phi) $: 
\begin{align*}
 \mathbb{\RN{2}}_{22}(\theta, \phi) 
 &= \begin{bmatrix}
     - R \sin \theta \cos\phi \\
     - R \sin \theta \sin\phi \\
     0 
\end{bmatrix} -  \Gamma_{22}^1 \frac{\partial f}{\partial x_1} - \Gamma_{22}^2 \frac{\partial f}{\partial x_2}  \\
 &= \begin{bmatrix}
     - R \sin \theta \cos\phi \\
     - R \sin \theta \sin\phi \\
     0 
\end{bmatrix} -  (-\sin \theta \cos \theta) \frac{\partial f}{\partial x_1} - 0. \frac{\partial f}{\partial x_2}  \\
 &= \begin{bmatrix}
     - R \sin \theta \cos\phi \\
     - R \sin \theta \sin\phi \\
     0 
\end{bmatrix} 
+ \sin \theta \cos \theta \begin{bmatrix}
 R \cos \theta \cos \phi \\
  R \cos \theta \sin \phi \\
  - R \sin \theta
\end{bmatrix} \\
& = R\sin\theta \begin{bmatrix}
 -\cos\phi + \cos^2\theta\cos\phi \\
 -\sin\phi+ \cos^2\theta\sin\phi  \\
 -\sin\theta\cos\theta 
\end{bmatrix}\\
& = R\sin\theta \begin{bmatrix}
 -\sin^2\theta\cos\phi \\
 -\sin^2\theta\sin\phi  \\
 -\sin\theta\cos\theta 
\end{bmatrix} \\
& = R\sin^2\theta \begin{bmatrix}
 -\sin\theta\cos\phi \\
 -\sin\theta\sin\phi  \\
 -\cos\theta 
\end{bmatrix}.
\end{align*}


The inverse of the Riemannian metric matrix is:
\begin{equation}
    g_S(\theta, \phi)^{-1} = \begin{bmatrix}
    \frac{1}{R^2} & 0 \\
    0 & \frac{1}{R^2 \sin^2 \theta}
    \end{bmatrix}.
\end{equation}

The mean curvature vector is then (omitting its zero components):
\begin{align*}
    H_S(\theta, \phi)
    &= 
    \frac{1}{2}\text{Tr}
        \mathbb{\RN{2}}_{p} \\
    &= 
    \frac{1}{2}g^{11}
    {\mathbb{\RN{2}}_{11}}(\theta, \phi) 
    + \frac{1}{2}g^{22} 
    {\mathbb{\RN{2}}_{22}}(\theta, \phi)\\ 
    &= \frac{1}{2R^2}
    \begin{bmatrix}
     - R \sin \theta \cos\phi \\
     - R \sin \theta \sin \phi \\
     - R \cos \theta
\end{bmatrix}\\
    & \qquad\qquad+ \frac{1}{2R^2\sin^2(\theta)}
    R\sin^2\theta \begin{bmatrix}
 -\sin\theta\cos\phi \\
 -\sin\theta\sin\phi  \\
 -\cos\theta
\end{bmatrix}\\
    &= \frac{1}{2R}
    \begin{bmatrix}
     - \sin \theta \cos\phi \\
     - \sin \theta \sin \phi \\
     - \cos \theta 
\end{bmatrix}
    + \frac{1}{2R}
   \begin{bmatrix}
 -\sin\theta\cos\phi \\
 -\sin\theta\sin\phi  \\
 -\cos\theta 
\end{bmatrix}\\
    &= \frac{1}{2R}
    \begin{bmatrix}
      -2\sin\theta\cos\phi \\
    -2\sin\theta\sin\phi \\
     - 2\cos \theta
\end{bmatrix}\\
    &= -\frac{1}{R}
    \begin{bmatrix}
      \sin\theta\cos\phi \\
    \sin\theta\sin\phi \\
     \cos \theta 
\end{bmatrix}.
\end{align*}
Its norm is: $\|H(\theta,\phi)\| = \frac{1}{R}$, which is the expected formula.
\end{proof}

\subsection{Mean Curvatures of the Torus}

\begin{example}[Mean curvatures of the torus immersed in $\mathbb{R}^N$]
We consider the torus $T$ obtained by rotating a circle of radius $b$ and center $(a, 0)$ around the axis $z$, and immersed in $\mathbb{R}^N$. The norms of its mean curvature vector is: 
\begin{equation}
 \|H_T(\theta, \phi)\| =  \frac{R+2r\cos\phi}{r(R+r\cos(\phi))} , \qquad \forall \theta, \phi \in \mathcal{S}^1 \times \mathcal{S}^1.
\end{equation}
\end{example}

\begin{proof}
We compute the mean curvature of a torus immersed in $\mathbb{R}^N$:
\begin{align*}
   f: \mathcal{S}^1 \times \mathcal{S}^1 &\mapsto \mathbb{R}^N \\
   \theta, \phi &\mapsto f(\theta, \phi) 
    = P. 
    \begin{bmatrix}
    c(\phi) \cos \theta \\
    c(\phi) \sin \theta \\
    r\sin \phi \\
    0 \\
    \vdots \\
    0
    \end{bmatrix} + t,
\end{align*}
where $c(\phi) = R + r \cos \phi$.

We compute the Hessian:
\begin{equation}
   \frac{\partial^2 f}{\partial x_i \partial x_j}(\theta, \phi)  
    = \begin{bmatrix}
     \frac{\partial^2f^1}{\partial x_i \partial x_j }(\theta, \phi) \\
     \frac{\partial^2f^2}{\partial x_i \partial x_j }(\theta, \phi) \\
     \frac{\partial^2f^3}{\partial x_i \partial x_j }(\theta, \phi) \\
     0 \\
     \vdots \\
     0
    \end{bmatrix},
\end{equation}
where we use the conventions $x_1 = \theta$ and $x_2 = \phi$. In what follows, for conciseness of the derivations, we do not write the components $\alpha$ of $f^\alpha$ for $\alpha = 4, ..., N$, as they only contribute terms equal to 0.

We get:
\begin{align*}
&\frac{\partial^2 f}{\partial \theta^2}(\theta, \phi) 
= \begin{bmatrix}
     - c(\phi)\cos\theta \\
     - c(\phi)\sin\theta \\
     0 
\end{bmatrix},\\
&\frac{\partial^2 f}{\partial \phi^2}(\theta, \phi) 
= \begin{bmatrix}
     -r \cos \phi\cos\theta \\
     -r \cos \phi\sin\theta \\
     -r\sin\phi
\end{bmatrix}.
\end{align*}

We only compute the diagonal terms, avoiding the computation of $\frac{\partial^2 f}{\partial \theta \partial \phi}(\theta, \phi)$ because we only need the diagonal terms in the definition of the trace, given that the inverse of the pullback metric is diagonal.

We compute the Hessian with respect to the pullback metric, again omitting its components for $\alpha > 3$.
\begin{equation}
    \mathbb{\RN{2}}_{ij}(\theta, \phi) 
    = \begin{bmatrix}
     \frac{\partial^2f^1}{\partial x_i \partial x_j }(\theta, \phi) - \sum_{k=1}^2 \Gamma_{ij}^k \frac{\partial f^1}{\partial x_k} \\
     \frac{\partial^2f^2}{\partial x_i \partial x_j }(\theta, \phi) - \sum_{k=1}^2 \Gamma_{ij}^k \frac{\partial f^2}{\partial x_k}  \\
     \frac{\partial^2f^3}{\partial x_i \partial x_j }(\theta, \phi) - \sum_{k=1}^2 \Gamma_{ij}^k \frac{\partial f^3}{\partial x_k}
    \end{bmatrix}.
\end{equation}

For the torus, the Christoffel symbols are: 
\begin{align*}
   &\Gamma_{11}^1=\Gamma_{22}^1=\Gamma_{22}^2=0,\\
   &\Gamma_{12}^1=\Gamma_{21}^1=- \frac{r\sin\phi}{c(\phi)},\\
  &\Gamma_{11}^2=\frac{1}{r}\sin\phi c(\phi),
\end{align*}
so that we get:
\begin{align*}
 \mathbb{\RN{2}}_{11}(\theta, \phi) 
 &= \begin{bmatrix}
     - c(\phi)\cos\theta \\
     - c(\phi)\sin\theta \\
     0
\end{bmatrix} 
-  \Gamma_{11}^1 \frac{\partial f}{\partial x_1} 
- \Gamma_{11}^2 \frac{\partial f}{\partial x_2}  \\
 &= \begin{bmatrix}
     - c(\phi)\cos\theta \\
     - c(\phi)\sin\theta \\
     0 
\end{bmatrix} 
- 0 \frac{\partial f}{\partial x_1} \\
&\qquad\qquad- \frac{1}{r}\sin\phi c(\phi) \begin{bmatrix}
-r\sin\phi \cos\theta \\
-r\sin\phi \sin\theta \\
r\cos\phi 
\end{bmatrix}  \\
  &= c(\phi) \begin{bmatrix}
 - \cos\theta +\sin^2\phi \cos\theta \\
 -\sin\theta +\sin^2\phi \sin\theta \\
 -\sin\phi\cos\phi 
  \end{bmatrix} \\
  &=  c(\phi) \begin{bmatrix}
 - \cos^2\phi \cos\theta \\
 -\cos^2\phi \sin\theta \\
 -\sin\phi\cos\phi 
  \end{bmatrix}\\
    &=  c(\phi)\cos\phi \begin{bmatrix}
 - \cos\phi \cos\theta \\
 -\cos\phi \sin\theta \\
 -\sin\phi 
  \end{bmatrix},
\end{align*}

and
\begin{align*}
 \mathbb{\RN{2}}_{22}(\theta, \phi) 
 &= \begin{bmatrix}
     -r \cos \phi\cos\theta \\
     -r \cos \phi\sin\theta \\
     -r\sin\phi
\end{bmatrix}
-  \Gamma_{22}^1 \frac{\partial f}{\partial x_1} 
- \Gamma_{22}^2 \frac{\partial f}{\partial x_2}  \\
 &= \begin{bmatrix}
     -r \cos \phi\cos\theta \\
     -r \cos \phi\sin\theta \\
     -r\sin\phi
\end{bmatrix}
- 0 \frac{\partial f}{\partial x_1} 
- 0 \frac{\partial f}{\partial x_2} \\
 &= -r \begin{bmatrix}
     \cos \phi\cos\theta \\
     \cos \phi\sin\theta \\
     \sin\phi
\end{bmatrix}.
\end{align*}

The inverse of the Riemannian metric matrix is:
\begin{equation}
    g_S(\theta, \phi)^{-1} = \begin{bmatrix}
    \frac{1}{(R+r\cos\phi)^2} & 0 \\
    0 & \frac{1}{r^2}
    \end{bmatrix}.
\end{equation}

The mean curvature vector is then:
\begin{align*}
    H_S(\theta, \phi)
    &= 
    \frac{1}{2}\text{Tr}
        \mathbb{\RN{2}}_{p} \\
    &= 
    \frac{1}{2}g^{11}
    {\mathbb{\RN{2}}_{11}}(\theta, \phi) 
    + \frac{1}{2}g^{22} 
    {\mathbb{\RN{2}}_{22}}(\theta, \phi)\\ 
    &= \frac{1}{2c^2(\phi)}
    c(\phi)\cos\phi \begin{bmatrix}
 - \cos\phi \cos\theta \\
 -\cos\phi \sin\theta \\
 -\sin\phi
  \end{bmatrix}\\
    &\qquad\qquad\qquad+ \frac{1}{2r^2}
    (-r) \begin{bmatrix}
     \cos \phi\cos\theta \\
     \cos \phi\sin\theta \\
     \sin\phi
\end{bmatrix}\\
    &= \frac{\cos\phi}{2c(\phi)}
   \begin{bmatrix}
 - \cos\phi \cos\theta \\
 -\cos\phi \sin\theta \\
 -\sin\phi 
  \end{bmatrix}
    - \frac{1}{2r}
    \begin{bmatrix}
     \cos \phi\cos\theta \\
     \cos \phi\sin\theta \\
     \sin\phi
\end{bmatrix}\\
    &= -\left(\frac{\cos\phi}{2c(\phi)}+\frac{1}{2r}\right)
    \begin{bmatrix}
 2\cos\phi \cos\theta \\
 2\cos\phi \sin\theta \\
 2\sin\phi
  \end{bmatrix}\\
    &= -\left(\frac{\cos\phi}{c(\phi)}+\frac{1}{r}\right)
    \begin{bmatrix}
 \cos\phi \cos\theta \\
 \cos\phi \sin\theta \\
 \sin\phi
  \end{bmatrix}\\
      &=- \frac{r\cos\phi+R+r\cos\phi}{rc(\phi)}
    \begin{bmatrix}
 \cos\phi \cos\theta \\
 \cos\phi \sin\theta \\
 \sin\phi 
  \end{bmatrix}\\
        &= -\frac{R+2r\cos\phi}{r(R+r\cos(\phi))}
    \begin{bmatrix}
 \cos\phi \cos\theta \\
 \cos\phi \sin\theta \\
 \sin\phi 
  \end{bmatrix}.
\end{align*}
Its norm is: $\|H(\theta,\phi)\| = \frac{R+2r\cos\phi}{r(R+r\cos(\phi))}$, which is the expected formula.
\end{proof}

\section{Invariance under Reparameterizations}

We give the proof for Lemma~\ref{lem:invariance-reparameterizations}.

\begin{lemma}[Invariance with respect to reparameterizations] 
The curvature profile is invariant under reparameterizations $f \rightarrow f \circ \varphi^{-1}$ of the neural manifold $\mathcal{M}$.
\end{lemma}

\begin{proof}
The distance between two points on the latent manifold is given by:
\begin{align*}
    \text{dist}(z_0,z) = \int_{0}^{1} d\tau \sqrt{\frac{d\gamma^{c}}{d\tau}\frac{d\gamma^{d}}{d\tau}g_{cd}(\gamma(\tau))}
\end{align*}
Where $\gamma$ is a geodesic and $g$ is the pullback metric induced by $f$. Consider a reparameterization of the latent space $\tilde{z} = \varphi(z)$. The distance between $\tilde{z_0}$ and $\tilde{z}$ is then given by 
\begin{align*}
    \text{dist}(\tilde{z_0},\tilde{z}) = \int_{0}^{1} d\tau \sqrt{\frac{d\tilde{\gamma}^{a}}{d\tau}\frac{d\tilde{\gamma}^{b}}{d\tau}\tilde{g}_{ab}(\tilde{\gamma}(\tau))}
\end{align*}
Which we can write as
\begin{align*}
    \text{dist}(\tilde{z_0},\tilde{z}) = \int_{0}^{1} d\tau \sqrt{\frac{d\tilde{\gamma}^{a}}{d\gamma^c}
    \frac{d\gamma^{c}}{d\tau}
    \frac{d\tilde{\gamma}^{b}}{d\gamma^{d}}
    \frac{d\gamma^{d}}{d\tau}
    \tilde{g}_{ab}(\tilde{\gamma}(\tau))}
\end{align*}
Via the metric tensor transformation law,
\begin{align*}
    g_{cd}(\gamma) = \frac{d\tilde{\gamma}^{a}}{d\gamma^c}\frac{d\tilde{\gamma}^{b}}{d\gamma^{d}}\tilde{g}_{ab}(\tilde{\gamma}(\tau))
\end{align*}
we conclude that $\text{dist}(\tilde{z_0},\tilde{z}) = \text{dist}(z_0,z)$ when the latent manifold is endowed with the pullback metric. Thus, if we consider the mean curvature vector $H(z) = H(\text{dist}(z_0,z))$ using a reference point $z_0 = \tilde{z}_0$, we obtain a reparameterization-invariant curvature profile on the latent manifold.
\end{proof}

\section{Examples of Latent Losses}

We illustrate the latent loss term presented in the main text with its explicit formulae for manifolds with topology $S^1$ or $S^2$.

\begin{example}[Latent loss terms for $\mathcal{Z}=\mathcal{S}^1$, $\mathcal{S}^2$]
The latent loss terms for neural manifolds parameterized by template manifolds $\mathcal{S}^1$, $\mathcal{S}^2$ are given by:
\begin{align*}
    \mathcal{L}^{\mathcal{S}^1}_{latent} &= (1 - \cos{(\theta_{gt} - \hat{\theta})})^2,\\
    \mathcal{L}^{\mathcal{S}^2}_{latent} &= (1 - \cos{(\theta_{gt} - \hat{\theta})} \\
    &\qquad\qquad + \sin(\theta_{gt})\sin(\hat{\theta})(1-\cos{(\phi_{gt} - \hat{\phi})}))^2.
\end{align*}
\end{example}

We implement these loss terms when we seek to enforce a canonical parameterization of the latent space, informed by outside world's task variables.


\section{Invariance under Permutations}

We give the proof for Lemma~\ref{lem:invariance-permutation}.

\begin{lemma}[Invariance of topology and geometry]
Consider a neural manifold $\mathcal{M}$ embedded in neural state space $\mathbb{R}_{+}^N$ corresponding to the recording of $N$ neurons. Permuting the order of the $N$ neurons: (i) leaves the topology of $\mathcal{M}$ invariant, (ii) leaves the geometry of $\mathcal{M}$ invariant.
\end{lemma}

\begin{proof}
Consider a neural manifold $\mathcal{M}$ embedded in neural state space $\mathbb{R}_{+}^N$ corresponding to the recording of $N$ neurons. Consider $S_N$ the group of permutations of the set $\{1, \dots, N\}$ labelling the $N$ neurons. The group $S_N$ acts on $\mathbb{R}_{+}^N$ by permutating of the order in which neurons are recorded, \textit{i.e.}, by permuting the axes of the space $\mathbb{R}_{+}^N$ as:
\begin{align*}
    S_N &\times \mathbb{R}_{+}^N &\mapsto &\mathbb{R}_{+}^N \\
    \sigma \ , &\quad (x_1, \dots, x_N) & \mapsto &(x_{\sigma(1)}, \dots, x_{\sigma(N)})
\end{align*}

Consider one permutation $\sigma \in S_N$. We show that the topology and geometry of $\mathcal{M}$ is invariant with respect to $\sigma$, by showing that $\sigma$ is a linear, thus continuous, isometry of $\mathbb{R}_{+}^N$.

By properties of permutation, $\sigma$ can be written as a product of transpositions $\tau_{t}$'s:
\vspace{-0.3cm}
\begin{equation}
\vspace{0.5cm}
    \sigma = \Pi_{t=1}^T \tau_t
    \vspace{-0.7cm}
\end{equation}
where the product is taken in the sense of the composition. A given transposition $\tau_t$ exchanges only two neurons. If each transposition leaves the topology of $\mathcal{M}$ invariant, then so does $\sigma$. Thus, we show that any transposition $\tau_{t}$ leaves the topology of $\mathcal{M}$ invariant.

Without loss of generality, we can prove it for a transposition $\tau_t$ exchanging neuron $1$ and neuron $2$, which will simplify the notations. The transposition $\tau_t =\tau_{12}$ exchanges neuron $1$ and $2$, which corresponds to exchanging the first two axes, axis $x_1$ and axis $x_2$ in $\mathbb{R}_{+}^N$, while keeping all other axes invariant. The action of this transposition corresponds to the symmetry of hyperplane $x_1 = x_2$ within $\mathbb{R}_{+}^N$, as:
\begin{align*}
    \mathbb{R}_{+}^N 
    &\mapsto \mathbb{R}_{+}^N \\
    x = (x_1, x_2, \dots , x_N) 
    &\mapsto (x_2, x_1, \dots , x_N) = T_{12}.x
\end{align*}
where $T_{12}$ is the permutation matrix:
\begin{equation*}
T_{12} = \begin{pmatrix}
0 & 1 & 0 & \cdots & 0 \\
1 & 0 & 0 & \cdots & 0 \\
0 & 0 & 1 & \cdots & 0 \\
\vdots & \vdots & \vdots & \ddots & \vdots \\
0 & 0 & 0 & \cdots & 1
\end{pmatrix}.
\end{equation*}
We see that the transposition $\tau_{12}$ of two axes is a linear map expressed by matrix $T_{12}$. As a linear map, $\tau_{12}$ is continuous. Any continuous map acting on a manifold preserves the topology of this manifold. Consequently, any transposition, and thus any permutation, preserves the topology of the neural manifold $\mathcal{M}$.

Additionally, the matrix $T_{12}$ is orthogonal as we can show that $T_{12}^T.T_{12} = I_N$ where $I_N$ is the identity matrix of shape $N \times N$. Consequently, $T_{12}$ is an isometry of $\mathbb{R}_{+}^N$ that preserves the geometry of $\mathcal{M}$ embedded in $\mathbb{R}^N$. Consequently, any transposition, and thus any permutation, preserves the geometry of the neural manifold $\mathcal{M}$ in the sense that $d(p, q) = d(\sigma *, \sigma q)$: the distances along the manifold $\mathcal{M}$ are invariant.
\end{proof}

\section{Synthetic Datasets}

We detail generation of the synthetic datasets in our experiments of distorted circles, spheres and tori.

\paragraph{Distorted Circle Datasets} The distorted circle datasets are created from the immersion $f^{\mathcal{S}^{1}}_{\text{synth}}$:
\begin{equation}\label{eq:decoders1} 
  \begin{aligned}
& \quad \mathcal{S}^1 \to \mathbb{R}^N \\
& \quad \theta_i \mapsto  \mathcal{R}*[A(\theta_i)(\cos{\theta_i}, \sin{\theta},0,...,0)] + \eta_i,
 \end{aligned}
\end{equation}
with $\theta_i$ uniformly distributed on $\mathcal{S}^1$ for $i=1,\ldots, n$. In Eq.~\ref{eq:decoders1},  $A(\theta) = 1 + \alpha [\exp{(-5(\theta - \pi /2)^{2})} + \exp{(-5(\theta - 3\pi /2)^{2})}] $ where the parameter $\alpha$ modulates the amplitude around the ring, creating extrinsic curvature in the vicinity of $\pi/2$ and $3\pi/2$.  

\paragraph{Distorted Sphere Datasets} The distorted 2-sphere datasets are created via the immersion $f^{\mathcal{S}^{2}}_{\text{synth}}$: 
\begin{equation}\label{eq:decoders2}
  \begin{aligned}
&\mathcal{S}^2 \to \mathbb{R}^N \\
&(\theta_i,\phi_i) \mapsto  \mathcal{R}*\big[A(\theta_i,\phi_i)\\
&\quad \quad \quad (\sin{\theta_i}\cos{\phi_i}, \sin{\theta_i}\sin{\phi_i},\cos{\theta_i},...,0)\big] + \eta_i,
 \end{aligned}
\end{equation}
with $\theta_i, \phi_i$ uniformly distributed on $\mathcal{S}^2$ for $i=1,\ldots, n$. In Eq.~\ref{eq:decoders2}, $A(\theta,\phi) = 1 + \alpha \exp{(-5(\theta)^{2})} + \alpha \exp{(-5(\theta - \pi /2)^{2})}$ with the parameter $\alpha$ introducing curvature in the vicinity of the north and south poles of $\mathcal{S}^2$.

\paragraph{Distorted Torus Datasets} The distorted 2-torus datasets are created via the immersion $f^{\mathcal{T}^{2}}_{\text{synth}}$:
\begin{equation}\label{eq:decodersT2}
  \begin{aligned} 
&\mathcal{T}^2 \to \mathbb{R}^N \\
&(\theta_i,\phi_i) \mapsto  \mathcal{R}*\big[A(\theta_i,\phi_i)\\
&\quad ((R-r\cos{\theta_i})\cos{\phi_i},(R-r\cos{\theta_i})\sin{\phi_i} ,r\sin{\theta_i},...,0)\big] \\
&\quad +\eta_i,
 \end{aligned}
\end{equation}
with $\theta_i, \phi_i$ uniformly distributed on $\mathcal{T}^2$ for $i=1,\ldots, n$. Here, $R$ and $r$ are the major and minor radii of the torus; these are assumed to carry no relevant information, and are both set to unity. The amplitude function $A(\theta,\phi)$ in Eq.~\ref{eq:decodersT2} is given by
\begin{align*}  
    A(\theta_i,\phi_i) &= 1 + \alpha \exp{(-2(\theta-\pi)^2)}[ \exp{(-2(\phi -\pi/2)^2)} \\
    &\qquad\qquad+ \exp{(-2(\phi -3\pi/2)^2)}] 
\end{align*}
with the parameter $\alpha$ introducing extrinsic curvature by stretching the torus on opposite sides at $(\theta,\phi) = (\pi,\pi/2)$ and $(\theta,\phi) = (\pi,3\pi/2)$

\begin{figure}
    \centering
    \includegraphics[width=0.75\linewidth]{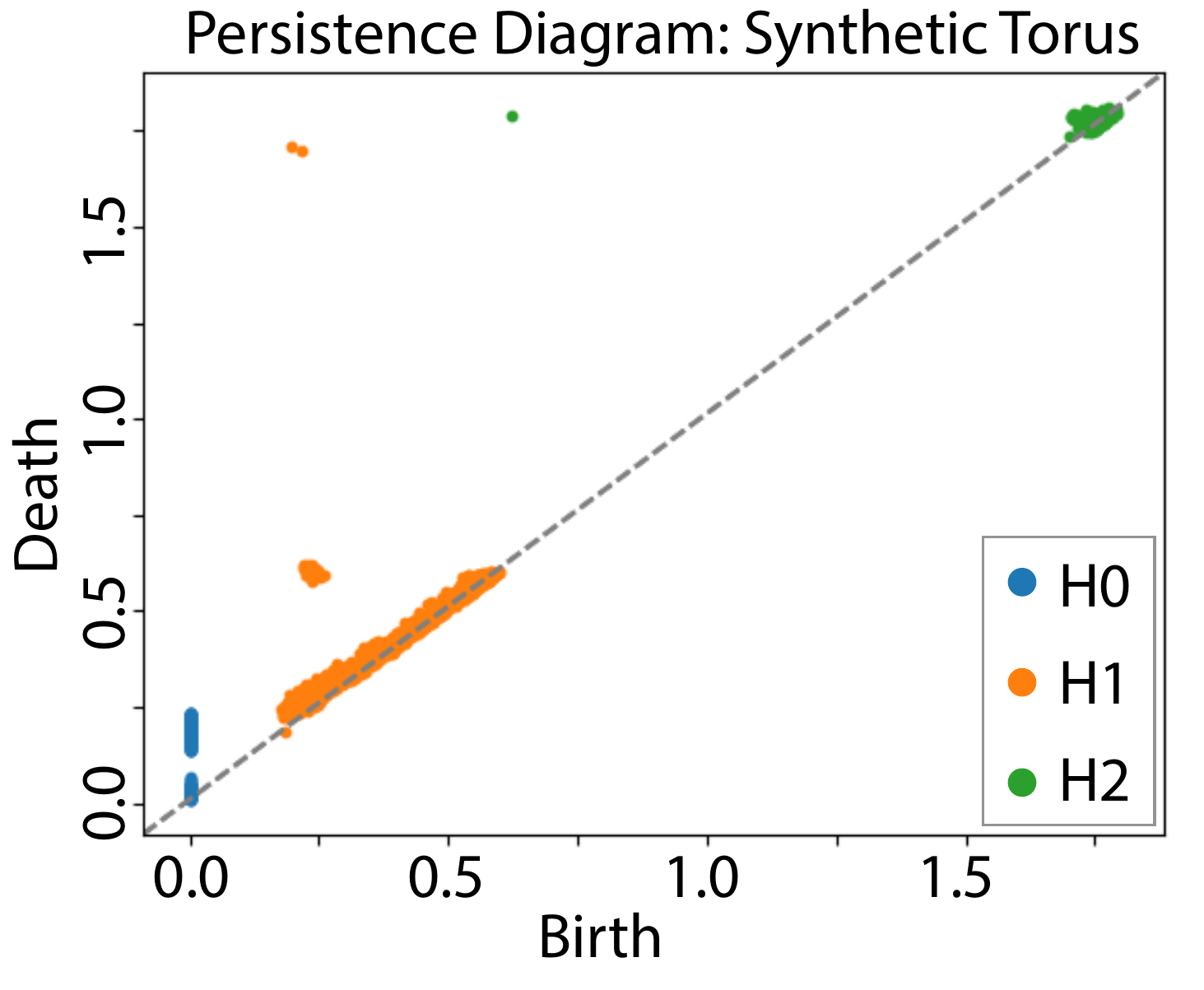}
    \caption{Persistence diagram for synthetic dataset on the torus, illustrating that TDA is an appropriate tool to compute the topology of a neural manifold and to constrain the latent space to be a given template $\mathcal{M}^*$.}
    \label{fig:learned-topo}
\end{figure}

\paragraph{Validation of Learned Topology} We show here the results of TDA applied to the synthetic dataset of the distorted $T^2$. This validates that the first step of our pipeline can effectively capture the topology of the manifold, as we observe the two holes know to characterize the torus topology in Fig.~\ref{fig:learned-topo}.

\begin{figure}[h!]
    \centering
    \includegraphics[width=1\linewidth]{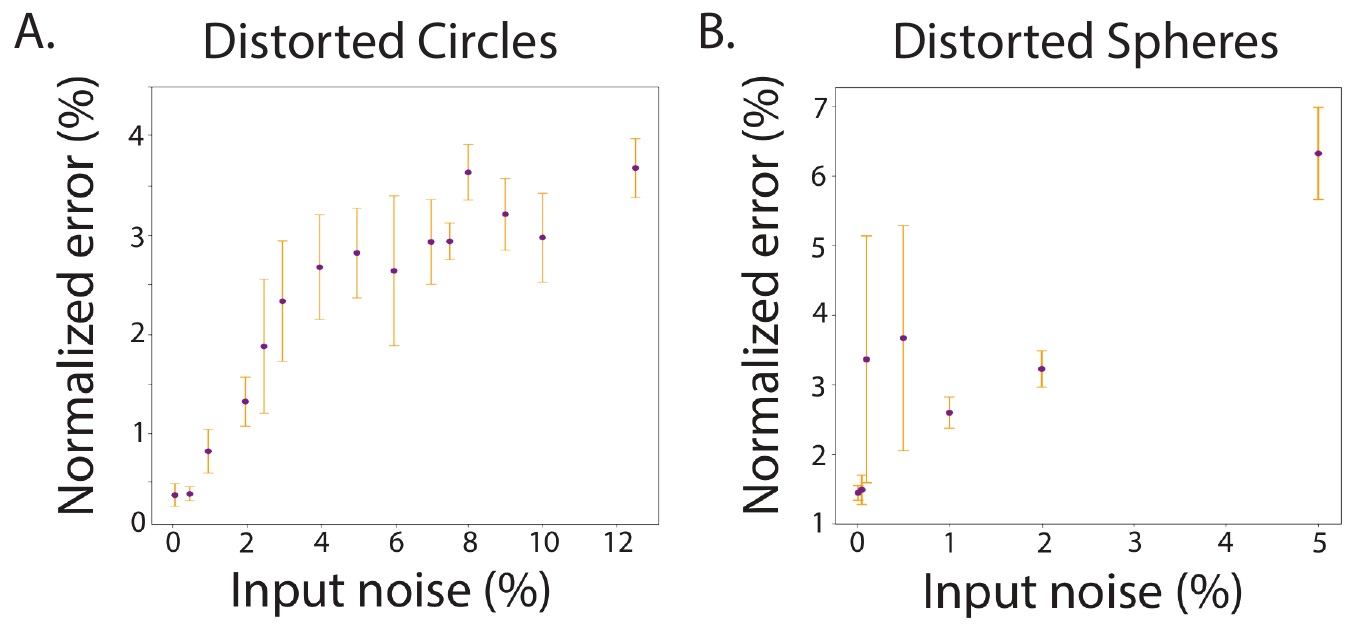}
    \caption{Curvature estimation error on distorted circles (A) and spheres (B). The number of neurons is fixed at $N=2$ for the distorted circles, and $N=3$ for the distorted spheres. Each experiment is repeated 5 times. The vertical orange bars show the -/+ 1 standard deviations of the curvature estimation errors.}
    \label{fig:error_n}
\end{figure}

\paragraph{Effect of Noise on Curvature Estimation Error for Distorted Spheres} We quantify the curvature estimation error as we vary the noise levels for distorted spheres, to complement the similar experiments presented in the main text for distorted circles. Fig.~\ref{fig:error_n} compares the curvature error for the circles (A) and the spheres (B).

\begin{figure}[h!]
    \centering
    \includegraphics[width=\linewidth]{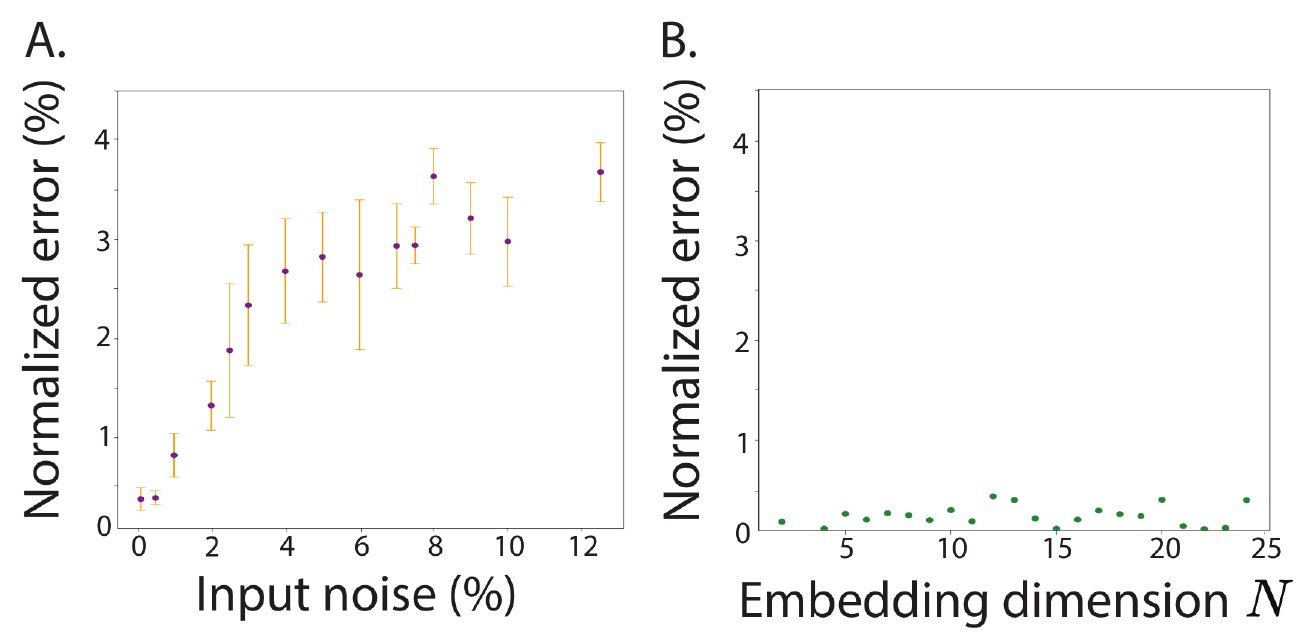}
    \caption{Curvature estimation error on distorted circles. A. While the error increases with the noise level, it does not go over 4\% for a range of noise levels corresponding to realistic values observed in neuroscience. Each experiment is repeated 5 times. The vertical orange bars show the -/+ 1 standard deviations of the errors. B. The error shows minimal variations with respect to the number of recorded neurons $N$. The vertical axis is shared across both plots for ease of comparison. }
    \label{fig:error_est}
\end{figure}

In these experiments, the number of neurons $N$ is fixed at $N=2$ for the circles and $N=3$ for the spheres. For each value of $\sigma$, 5 synthetic manifolds are generated and estimated. The vertical orange bars represent +/- standard deviation. We observe that the error is approximately twice as important in the case of the spheres than in the case of the circles. While the results in the main text seem to indicate that the estimation error does not depend on the number of neurons, we could conjecture that it depends linearly in the dimension of the manifold.

\section{Experimental Place Cells (12 neurons)}

We used data from 12 place cells within one session, whose neural spikes are binned with time-steps of 1 second, to yield 828 time points of neural activity in $\mathbb{R}_{+}^{12}$. Our results show that the reconstructed activations match the recorded (ground-truth), see Fig.~\ref{fig:expt41}~(A): even if we cannot observe the neural manifold in $\mathbb{R}_{+}^{12}$, we expect it to be correctly reconstructed. The canonical parameterization is correctly learned in the VAE latent space, as shown in Fig.~\ref{fig:expt41}~(B). The curvature profile is shown in Fig.~\ref{fig:expt41}~(C) where the angle is the physical lab angle. As for the simulated place cells, we observe several peaks which correspond to the place fields of the neurons: e.g. neuron 4 shown in Fig.~\ref{fig:expt41}~(C) which codes for one of the largest peaks, which is expected as it has the strongest activation from Fig.~\ref{fig:expt41}~(A). We reproduce this experiment on another dataset with 40 place cells recorded from another animal and find similar results in the supplementary materials. We emphasize that the goal of this experiment is not to reveal new neuroscience insights, but rather to show how the results provided by the curvature profiles are consistent with what one would expect and with what neuroscientists already know.

\section{Experimental Place Cells (40 neurons)}\label{app:expt_place_cells}

We perform the same experiment on real place cell data as in subsection~\ref{sec:expt_place_cells}, this time recording from 40 neurons. In this experiment, after the temporal binning, we have 8327 time points. Similarly to the previous experiment, the reconstruction of the neural activity together with the canonical parameterization of the latent space are correctly learned by the model. As expected, we observe a neural manifold whose geometry shows more ``petals'' which intuitively correspond to the higher number of neurons recorded by this experiment. We locate a place cell whose place field provides one of the highest peaks, neuron 22, and color the curvature profile based on the activity of this neuron. 

\begin{figure}
    \centering
    \vspace{-0.2cm}
    \includegraphics[width=1\linewidth]{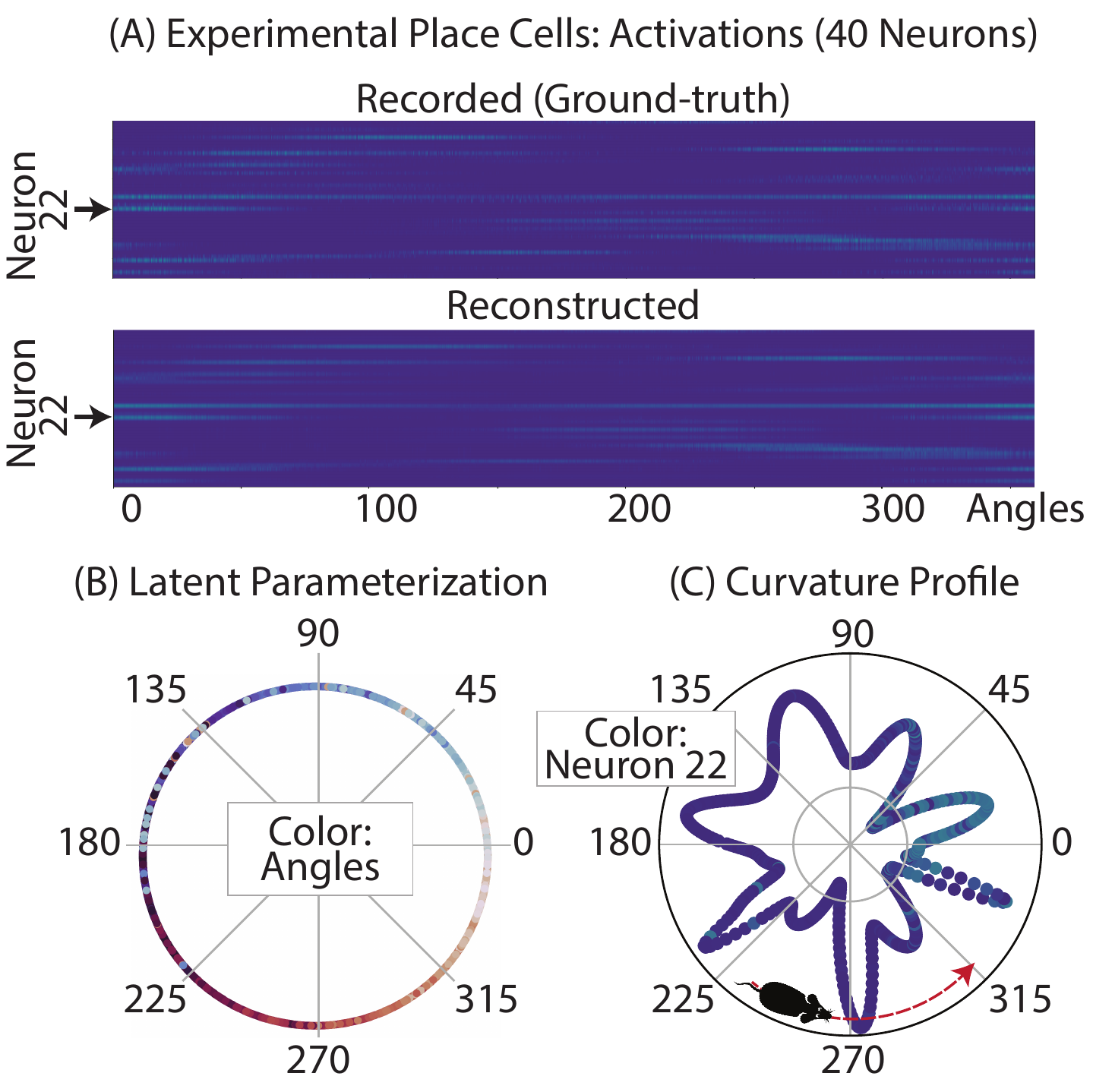}
    \caption{Neural geometry of 40 experimental place cells as an animal moves along a circle. (A) Recorded versus reconstructed neural activity of 12 place cells with respect to the positional angles of the animal in lab space. (B) Latent space's parameterization: the angular latent variables are colored by the corresponding positional angles of the animal in lab space. (C) Curvature profile of the neural manifold in log scale: the angles represent the physical lab space angles, colored by the reconstructed activation of neuron 22.}
    \label{fig:expt34}
\end{figure}

This demonstrates that our method can be applied to real neural datasets, providing geometric results that match the intuition. In the case of hippocampal place cells, the geometry of the neural manifold depends on the number of neurons, whether they tile the physical space where the animal is moving. It also depends on the spiking profile of their place fields, including its amplitude and width.

%% file: main.bbl
\begin{thebibliography}{10}\itemsep=-1pt

\bibitem{aronov2017mapping}
Dmitriy Aronov, Rhino Nevers, and David~W. Tank.
\newblock Mapping of a non-spatial dimension by the hippocampal–entorhinal
  circuit.
\newblock {\em Nature}, 543(7647):719--722, 2017.
\newblock arXiv: 10.1038/nature21692 Publisher: Nature Publishing Group.

\bibitem{arvanitidis2017latent}
Georgios Arvanitidis, Lars~Kai Hansen, and S{\o}ren Hauberg.
\newblock Latent space oddity: on the curvature of deep generative models.
\newblock {\em arXiv preprint arXiv:1710.11379}, 2017.

\bibitem{aubin1998some}
Thierry Aubin.
\newblock {\em Some nonlinear problems in Riemannian geometry}.
\newblock Springer Science \& Business Media, 1998.

\bibitem{bengio2013representation}
Yoshua Bengio, Aaron Courville, and Pascal Vincent.
\newblock Representation learning: A review and new perspectives.
\newblock {\em IEEE transactions on pattern analysis and machine intelligence},
  35(8):1798--1828, 2013.

\bibitem{bjerke2022understanding}
Martin Bjerke, Lukas Schott, Kristopher~T Jensen, Claudia Battistin, David~A
  Klindt, and Benjamin~A Dunn.
\newblock Understanding neural coding on latent manifolds by sharing features
  and dividing ensembles.
\newblock {\em arXiv preprint arXiv:2210.03155}, 2022.

\bibitem{blei2017variational}
David~M. Blei, Alp Kucukelbir, and Jon~D. McAuliffe.
\newblock Variational {Inference}: {A} {Review} for {Statisticians}.
\newblock {\em Journal of the American Statistical Association},
  112(518):859--877, Apr. 2017.
\newblock arXiv:1601.00670 [cs, stat].

\bibitem{boccara2019entorhinal}
Charlotte~N Boccara, Michele Nardin, Federico Stella, Joseph O’Neill, and
  Jozsef Csicsvari.
\newblock The entorhinal cognitive map is attracted to goals.
\newblock {\em Science}, 363(6434):1443--1447, 2019.

\bibitem{chadebec2022geometric}
Cl{\'e}ment Chadebec and St{\'e}phanie Allassonni{\`e}re.
\newblock A geometric perspective on variational autoencoders.
\newblock {\em arXiv preprint arXiv:2209.07370}, 2022.

\bibitem{chadebec2020geometry}
Cl{\'e}ment Chadebec, Cl{\'e}ment Mantoux, and St{\'e}phanie Allassonni{\`e}re.
\newblock Geometry-aware hamiltonian variational auto-encoder.
\newblock {\em arXiv preprint arXiv:2010.11518}, 2020.

\bibitem{chaudhuri2019intrinsic}
Rishidev Chaudhuri, Berk Ger{\c{c}}ek, Biraj Pandey, Adrien Peyrache, and Ila
  Fiete.
\newblock The intrinsic attractor manifold and population dynamics of a
  canonical cognitive circuit across waking and sleep.
\newblock {\em Nature neuroscience}, 22(9):1512--1520, 2019.

\bibitem{chen2018metrics}
Nutan Chen, Alexej Klushyn, Richard Kurle, Xueyan Jiang, Justin Bayer, and
  Patrick Smagt.
\newblock Metrics for deep generative models.
\newblock In {\em International Conference on Artificial Intelligence and
  Statistics}, pages 1540--1550. PMLR, 2018.

\bibitem{chung2021neural}
SueYeon Chung and LF Abbott.
\newblock Neural population geometry: An approach for understanding biological
  and artificial neural networks.
\newblock {\em Current opinion in neurobiology}, 70:137--144, 2021.

\bibitem{connor2021variational}
Marissa Connor, Gregory Canal, and Christopher Rozell.
\newblock Variational autoencoder with learned latent structure.
\newblock In {\em International Conference on Artificial Intelligence and
  Statistics}, pages 2359--2367. PMLR, 2021.

\bibitem{curto2017can}
Carina Curto.
\newblock What can topology tell us about the neural code?
\newblock {\em Bulletin of the American Mathematical Society}, 54(1):63--78,
  2017.

\bibitem{danjo2018spatial}
Teruko Danjo, Taro Toyoizumi, and Shigeyoshi Fujisawa.
\newblock Spatial representations of self and other in the hippocampus.
\newblock {\em Science}, 359(6372):213--218, 2018.

\bibitem{davidson2018hyperspherical}
Tim~R Davidson, Luca Falorsi, Nicola De~Cao, Thomas Kipf, and Jakub~M Tomczak.
\newblock Hyperspherical variational auto-encoders.
\newblock {\em arXiv preprint arXiv:1804.00891}, 2018.

\bibitem{dicarlo2007untangling}
James~J DiCarlo and David~D Cox.
\newblock Untangling invariant object recognition.
\newblock {\em Trends in cognitive sciences}, 11(8):333--341, 2007.

\bibitem{doersch2021tutorial}
Carl Doersch.
\newblock Tutorial on {Variational} {Autoencoders}.
\newblock Technical Report arXiv:1606.05908, arXiv, Jan. 2021.
\newblock arXiv:1606.05908 [cs, stat] type: article.

\bibitem{falorsi2018explorations}
Luca Falorsi, Pim De~Haan, Tim~R Davidson, Nicola De~Cao, Maurice Weiler,
  Patrick Forr{\'e}, and Taco~S Cohen.
\newblock Explorations in homeomorphic variational auto-encoding.
\newblock {\em arXiv preprint arXiv:1807.04689}, 2018.

\bibitem{gao2015simplicity}
Peiran Gao and Surya Ganguli.
\newblock On simplicity and complexity in the brave new world of large-scale
  neuroscience.
\newblock {\em Current opinion in neurobiology}, 32:148--155, 2015.

\bibitem{gardner2022toroidal}
Richard~J Gardner, Erik Hermansen, Marius Pachitariu, Yoram Burak, Nils~A Baas,
  Benjamin~A Dunn, May-Britt Moser, and Edvard~I Moser.
\newblock Toroidal topology of population activity in grid cells.
\newblock {\em Nature}, 602(7895):123--128, 2022.

\bibitem{ghrist2007barcodes}
Robert Ghrist.
\newblock Barcodes: {The} persistent topology of data.
\newblock {\em Bulletin of the American Mathematical Society}, 45(01):61--76,
  Oct. 2007.

\bibitem{hafting2005microstructure}
Torkel Hafting, Marianne Fyhn, Sturla Molden, May-Britt Moser, and Edvard~I.
  Moser.
\newblock Microstructure of a spatial map in the entorhinal cortex.
\newblock 436(7052):801--806.

\bibitem{hauberg2019bayes}
Søren Hauberg.
\newblock Only {{Bayes}} should learn a manifold (on the estimation of
  differential geometric structure from data).

\bibitem{hauser2017principles}
Michael Hauser and Asok Ray.
\newblock Principles of riemannian geometry in neural networks.
\newblock {\em Advances in neural information processing systems}, 30, 2017.

\bibitem{Hermansen2022}
Erik Hermansen, David~A. Klindt, and Benjamin~A. Dunn.
\newblock Uncovering 2-d toroidal representations in grid cell ensemble
  activity during 1-d behavior.
\newblock Nov. 2022.

\bibitem{jayakumar2019recalibration}
Ravikrishnan~P Jayakumar, Manu~S Madhav, Francesco Savelli, Hugh~T Blair,
  Noah~J Cowan, and James~J Knierim.
\newblock Recalibration of path integration in hippocampal place cells.
\newblock {\em Nature}, 566(7745):533--537, 2019.

\bibitem{kalatzis2020variational}
Dimitris Kalatzis, David Eklund, Georgios Arvanitidis, and S{\o}ren Hauberg.
\newblock Variational autoencoders with riemannian brownian motion priors.
\newblock {\em arXiv preprint arXiv:2002.05227}, 2020.

\bibitem{keinath2018environmental}
Alexandra~T Keinath, Russell~A Epstein, and Vijay Balasubramanian.
\newblock Environmental deformations dynamically shift the grid cell spatial
  metric.
\newblock {\em eLife}, 7:e38169, Oct. 2018.
\newblock Publisher: eLife Sciences Publications, Ltd.

\bibitem{kingma2013auto}
Diederik~P Kingma and Max Welling.
\newblock Auto-encoding variational bayes.
\newblock {\em arXiv preprint arXiv:1312.6114}, 2013.

\bibitem{knudsen2021hippocampal}
Eric~B. Knudsen and Joni~D. Wallis.
\newblock Hippocampal neurons construct a map of an abstract value space.
\newblock {\em Cell}, pages 1--11, 2021.
\newblock Publisher: Elsevier Inc.

\bibitem{kuhnel2018latent}
Line Kuhnel, Tom Fletcher, Sarang Joshi, and Stefan Sommer.
\newblock Latent space non-linear statistics.
\newblock {\em arXiv preprint arXiv:1805.07632}, 2018.

\bibitem{madhav2022dome}
Manu~S Madhav, Ravikrishnan~P Jayakumar, Shahin~G Lashkari, Francesco Savelli,
  Hugh~T Blair, James~J Knierim, and Noah~J Cowan.
\newblock The dome: A virtual reality apparatus for freely locomoting rodents.
\newblock {\em Journal of Neuroscience Methods}, 368:109336, 2022.

\bibitem{madhav2022closed-loop}
Manu~S. Madhav, Ravikrishnan~P. Jayakumar, Brian Li, Francesco Savelli,
  James~J. Knierim, and Noah~J. Cowan.
\newblock Closed-loop control and recalibration of place cells by optic flow.
\newblock {\em bioRxiv}, 2022.

\bibitem{mathieu2019continuous}
Emile Mathieu, Charline Le~Lan, Chris~J Maddison, Ryota Tomioka, and Yee~Whye
  Teh.
\newblock Continuous hierarchical representations with poincar{\'e} variational
  auto-encoders.
\newblock {\em Advances in neural information processing systems}, 32, 2019.

\bibitem{mikulski2019toroidal}
Maciej Mikulski and Jaroslaw Duda.
\newblock Toroidal autoencoder.
\newblock {\em arXiv preprint arXiv:1903.12286}, 2019.

\bibitem{miolane2020geomstats}
Nina Miolane, Nicolas Guigui, Alice~Le Brigant, Johan Mathe, Benjamin Hou, Yann
  Thanwerdas, Stefan Heyder, Olivier Peltre, Niklas Koep, Hadi Zaatiti, Hatem
  Hajri, Yann Cabanes, Thomas Gerald, Paul Chauchat, Christian Shewmake, Daniel
  Brooks, Bernhard Kainz, Claire Donnat, Susan Holmes, and Xavier Pennec.
\newblock Geomstats: A python package for riemannian geometry in machine
  learning.
\newblock {\em Journal of Machine Learning Research}, 21(223):1--9, 2020.

\bibitem{moser2008place}
Edvard~I Moser, Emilio Kropff, and May-Britt Moser.
\newblock Place cells, grid cells, and the brain's spatial representation
  system.
\newblock {\em Annu. Rev. Neurosci.}, 31:69--89, 2008.

\bibitem{nieh2021geometry}
Edward~H. Nieh, Manuel Schottdorf, Nicolas~W. Freeman, Ryan~J. Low, Sam
  Lewallen, Sue~Ann Koay, Lucas Pinto, Jeffrey~L. Gauthier, Carlos~D. Brody,
  and David~W. Tank.
\newblock Geometry of abstract learned knowledge in the hippocampus.
\newblock {\em Nature}, (February 2020), 2021.
\newblock Publisher: Springer US.

\bibitem{okeefe1971hippocampus}
John O'Keefe and Jonathan Dostrovsky.
\newblock The hippocampus as a spatial map: preliminary evidence from unit
  activity in the freely-moving rat.
\newblock {\em Brain research}, 1971.

\bibitem{omer2018social}
David~B Omer, Shir~R Maimon, Liora Las, and Nachum Ulanovsky.
\newblock Social place-cells in the bat hippocampus.
\newblock {\em Science (New York, N.Y.)}, 359(6372):218--224, 2018.

\bibitem{pearson1901lines}
Karl Pearson.
\newblock {LIII. On lines and planes of closest fit to systems of points in
  space}, Nov. 1901.

\bibitem{roweis2000nonlinear}
Sam~T. Roweis and Lawrence~K. Saul.
\newblock Nonlinear dimensionality reduction by locally linear embedding.
\newblock {\em Science}, 290(5500):2323–2326, 2000.

\bibitem{savelli2017framing}
Francesco Savelli, J.~D. Luck, and James~J. Knierim.
\newblock Framing of grid cells within and beyond navigation boundaries.
\newblock {\em eLife}, 6:1--29, 2017.

\bibitem{shao2018riemannian}
Hang Shao, Abhishek Kumar, and P Thomas~Fletcher.
\newblock The riemannian geometry of deep generative models.
\newblock In {\em Proceedings of the IEEE Conference on Computer Vision and
  Pattern Recognition Workshops}, pages 315--323, 2018.

\bibitem{tenenbaum2000global}
Joshua~B. Tenenbaum, Vin~de Silva, and John~C. Langford.
\newblock A global geometric framework for nonlinear dimensionality reduction.
\newblock {\em Science}, 290(5500):2319–2323, 2000.

\bibitem{vandermaaten2008visualizing}
Laurens van~der Maaten and Geoffrey Hinton.
\newblock Visualizing data using t-sne.
\newblock {\em Journal of Machine Learning Research}, 9:2579--2605, 11 2008.

\bibitem{vyas2020computation}
Saurabh Vyas, Matthew~D Golub, David Sussillo, and Krishna~V Shenoy.
\newblock Computation through neural population dynamics.
\newblock {\em Annual Review of Neuroscience}, 43:249, 2020.

\bibitem{wattenberg2016use}
Martin Wattenberg, Fernanda Vi{\'e}gas, and Ian Johnson.
\newblock How to use t-sne effectively.
\newblock {\em Distill}, 1(10):e2, 2016.

\end{thebibliography}
